\documentclass[a4paper,USenglish,cleveref, autoref]{lipics-v2019}
\usepackage{amsmath, amssymb, amsthm, amsfonts,alltt,clrscode}
\usepackage{array,colortbl,hhline,xcolor}
\usepackage{graphicx}
\usepackage[labelfont=bf,font=small]{caption}
\usepackage{url}
\usepackage{float}
\usepackage[list=true,listformat=simple]{subcaption}

\nolinenumbers
\usepackage{hyperref}

\def\emph#1{\textbf{\textit{\boldmath #1}}}

\graphicspath{{./figures/simulate_l2t/}{./figures/planarity/}{./figures/NCL/}{./figures/locking_2-toggle/}{./figures/2-toggle-paper/}{./figures/}}

{\makeatletter
 \gdef\xxxmark{%
   \expandafter\ifx\csname @mpargs\endcsname\relax 
     \expandafter\ifx\csname @captype\endcsname\relax 
       \marginpar{xxx}
     \else
       xxx 
     \fi
   \else
     xxx 
   \fi}
 \gdef\xxx{\@ifnextchar[\xxx@lab\xxx@nolab}
 \long\gdef\xxx@lab[#1]#2{\textbf{[\xxxmark #2 ---{\sc #1}]}}
 \long\gdef\xxx@nolab#1{\textbf{[\xxxmark #1]}}
}

\title{Toward a General Complexity Theory of Motion Planning: 
  Characterizing Which Gadgets Make Games Hard}

\titlerunning{Toward a General Complexity Theory of Motion Planning}

\author{Erik D. Demaine}{MIT Computer Science and Artificial Intelligence Laboratory, 32 Vassar Street, Cambridge, MA 02139, USA}{edemaine@mit.edu}{}{}
\author{Dylan H. Hendrickson}{MIT Computer Science and Artificial Intelligence Laboratory, 32 Vassar Street, Cambridge, MA 02139, USA}{dylanhen@mit.edu}{https://orcid.org/0000-0002-9967-8799
}{}
\author{Jayson Lynch}{MIT Computer Science and Artificial Intelligence Laboratory, 32 Vassar Street, Cambridge, MA 02139, USA}{jaysonl@mit.edu}{https://orcid.org/0000-0003-0801-1671}{}

\authorrunning{E. D. Demaine, D. H. Hendrickson, and J. Lynch}

\Copyright{E. D. Demaine, D. H. Hendrickson, and J. Lynch}



\acknowledgements{This work grew out of an open problem session from MIT class on
Algorithmic Lower Bounds: Fun with Hardness Proofs (6.890) from Fall 2014.}

\EventEditors{Thomas Vidick}
\EventNoEds{1}
\EventLongTitle{11th Innovations in Theoretical Computer Science Conference (ITCS 2020)}
\EventShortTitle{ITCS 2020}
\EventAcronym{ITCS}
\EventYear{2020}
\EventDate{January 12--14, 2020}
\EventLocation{Seattle, Washington, USA}
\EventLogo{}
\SeriesVolume{151}
\ArticleNo{62}
\begin{document}

\maketitle

\begin{abstract}
We begin a general theory for characterizing the computational complexity of
motion planning of robot(s) through a graph of ``gadgets,'' where each gadget
has its own state defining a set of allowed traversals which in turn modify
the gadget's state.
We study two general families of such gadgets within this theory,
one which naturally leads to motion planning problems with polynomially bounded
solutions, and another which leads to polynomially unbounded (potentially
exponential) solutions.
We also study a range of competitive game-theoretic scenarios,
from one player controlling one robot to teams of players each controlling
their own robot and racing to achieve their team's goal.
Under certain restrictions on these gadgets, we
fully characterize the complexity of
bounded 1-player motion planning (NL vs.\ NP-complete),
unbounded 1-player motion planning (NL vs.\ PSPACE-complete),
and bounded 2-player motion planning (P vs.\ PSPACE-complete),
and we partially characterize the complexity of
unbounded 2-player motion planning (P vs.\ EXPTIME-complete),
bounded 2-team motion planning (P vs.\ NEXPTIME-complete), and
unbounded 2-team motion planning (P vs.\ undecidable).
These results can be seen as an alternative to Constraint Logic
(which has already proved useful as a basis for hardness reductions),
providing a wide variety of agent-based gadgets,
any one of which suffices to prove a problem hard.
\end{abstract}

\section{Introduction}
\label{sec:intro}

Most hardness proofs are based on \emph{gadgets} --- local fragments,
each often representing corresponding fragments of the input instance,
that combine to form the overall reduction.
Garey and Johnson \cite{NPBook} called gadgets ``basic units'' and
the overall technique ``local replacement proofs''.
The search for a hardness reduction usually starts by experimenting with small
candidate gadgets, seeing how they behave, and repeating until amassing
a sufficient collection of gadgets to prove hardness.

This approach leads to a natural question: what gadget sets suffice to
prove hardness?  There are many possible answers to this question, depending
on the precise meaning of ``gadget'' and the style of problem considered.
Schaefer \cite{Schaefer-1978-SAT} characterized the complexity of all
\emph{Boolean constraint satisfiability} gadgets, with a dichotomy between
polynomial problems (e.g., 2SAT, Horn SAT, dual-Horn SAT, XOR SAT) and
NP-complete problems (e.g., 3SAT, 1-in-3SAT, NAE 3SAT).
At STOC'97, Khanna, Sudan, Trevisan, and Williamson~\cite{CSP-characterization}
refined this result to characterize \emph{approximability} of constraint
satisfaction problems, forking into polynomial, APX-complete, Poly-APX-complete,
Nearest-Codeword-complete, and Min-Horn-Deletion-complete.
Introduced at CCC'08, \emph{Constraint Logic} \cite{CL_Complexity2008,GPCBook09}
proves sufficiency of small
sets of gadgets on directed graphs that always satisfy one local rule
(weighted in-degree at least~$2$), in many game types
(1-player, 2-player, and team games, both polynomially bounded and unbounded),
although the exact minimal sets of required gadgets remain unknown.

The aforementioned general techniques naturally model ``global'' moves that
can be made anywhere at any time (while satisfying the constraints).
Nonetheless, the techniques have been successful at proving hardness for
problems where moves must be made local to an agent/robot that traverses the
instance.
For single-player agent-based problems, the \emph{doors-and-buttons} framework
(described in \cite{Forisek10} and improved by \cite{HardGames12} and \cite{van2015pspace}) is a good example of classifying a universe of abstract motion planning problems which can then be applied.
In addition, the \emph{door gadget} used to prove
Lemmings \cite{viglietta2015lemmings} and various Nintendo games \cite{Nintendo_TCS} PSPACE-complete served as a primary example of the form of gadget we wanted to generalize.

In this paper, we analyze which gadgets suffice for hardness in a general \emph{semi-static}
\emph{motion planning problem} where one or more agents/robots traverse a given
environment, which only changes in response to the agent's actions, from given start location(s) to given goal location(s).  We study
a very general model of gadget, where the gadget changes state when it gets
traversed by an agent according to a general transition function, enabling
and/or disabling certain traversals in the future.  We study this model
from the traditional single-robot (one-player) perspective, extending our
initial work on this case \cite{Toggles_FUN2018}, as well as from the
perspective of two robots or teams of robots competing to reach their
respective goals.  We also analyze natural settings where the number of moves
is polynomially bounded, because each gadget can be traversed only a bounded
number of times, or more general settings where gadgets can be re-used many
times and thus the number of moves can be exponential in the environment
complexity.  In each case, we partially or fully characterize which gadgets
suffice to make the motion planning problem hard (NP-hard, PSPACE-hard,
EXPTIME-hard, NEXPTIME-hard, or RE-hard, depending on the scenario),
and conversely which gadgets result in a polynomially solvable problem
(NL or P).
Table~\ref{results} summarizes our results.

\begin{table}
  \centering
  \newcolumntype{P}[1]{>{\raggedright\parskip=8pt}p{#1}}
  \def\arraystretch{1.4}
  \def\class#1{\textbf{#1}}
  \def\vs{~\textsc{vs.}~}
  \definecolor{purple}{rgb}{0.47,0.25,0.55}
  \def\HEADER{\cellcolor{gray!30}}
  \arrayrulecolor{purple!50!black}
  \setlength\arrayrulewidth{1pt}
  \def\REF#1{[\S\ref{#1}]}
  \caption{Summary of our results for $k$-tunnel gadgets (with additional
    constraints listed in the left column).
    A ``full characterization'' means that we
    give an easily checkable condition on the allowed gadget set that
    determines the complexity of the corresponding motion planning problem;
    a ``partial characterization'' means that we give two easily checkable
    conditions on the allowed gadget set, one for the easy class and one for
    the hard class, each of which suffices to establish the complexity
    of the corresponding motion planning problem.}
  \begin{tabular}{|P{6.5em}|P{9em}|P{9em}|P{9.5em}|}
    \hhline{~|-|-|-|}
      \multicolumn{1}{c|}{}
    & \multicolumn{1}{c|}{\HEADER{\textbf{1-Player Game}}}
    & \multicolumn{1}{c|}{\HEADER{\textbf{2-Player Game}}}
    & \multicolumn{1}{c|}{\HEADER{\textbf{Team Game}}}
    \cr
    \hline
    \HEADER{\textbf{Polynomially Bounded} \newline (DAG gadgets)\!\!}
    & \class{NL}\vs\class{NP-complete}: full characterization \REF{sec:1-Player Bounded}
    & \class{P}\vs\class{PSPACE-complete}: full characterization \REF{sec:2-Player Bounded}
    & \class{P}\vs\class{NEXPTIME-complete}: full characterization \REF{sec:Team Bounded}
    \cr
    \hline
    \HEADER{\textbf{Polynomially Unbounded} \newline (reversible, deterministic gadgets)}
    & \class{NL}\vs\class{PSPACE-complete}: full characterization \REF{sec:1-Player Unbounded}
      \\ 
      \par \class{Planar}: equivalent \REF{sec:1-Player Unbounded Planar}
    & \class{P}\vs\class{EXPTIME-complete}: partial characterization \REF{sec:2-Player Unbounded}
    & \class{P}\vs\class{RE-complete ($\Rightarrow$ Undecidable)}: partial characterization \REF{sec:Team Unbounded}
    \cr
    \hline
  \end{tabular}
  \label{results}
\end{table}

\subsection{Gadget Model and Motion-Planning Games}

\begin{figure}
  \centering
  \begin{subfigure}{0.5\textwidth}
    \centering
    \includegraphics[scale=0.75]{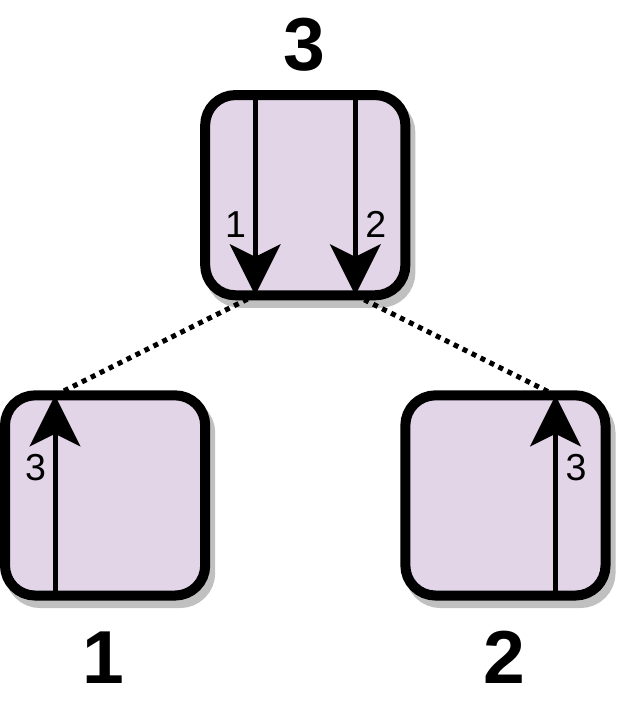}
    \caption{The locking 2-toggle gadget (L2T). In the top state 3, you can traverse either tunnel going down, which blocks off the other tunnel until you reverse the initial traversal.}
    \label{fig:L2T}
  \end{subfigure}\hfil\hfil
  \begin{subfigure}{0.4\textwidth}
    \centering
    \includegraphics[scale=0.75]{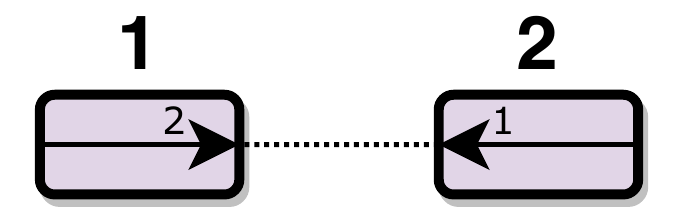}
    \caption{The 1-toggle gadget. Traversing the tunnel reverses the direction that it can be traversed.}
    \label{fig:1-toggle}
  \end{subfigure}
  \caption{Basic gadgets that can be simulated by any interacting-$k$-tunnel reversible deterministic gadget, as shown in Section~\ref{sec:reducing to locking 2-toggles}.}
  \label{fig:basic gadgets}
\end{figure}

In general, we model a \emph{gadget} as consisting of a finite number of
\emph{locations} (entrances/exits) and a finite number of \emph{states};
see Figure~\ref{fig:basic gadgets} for two examples.
We may also consider a family of gadgets parameterized by the problem size. 
In this case we restrict the number of locations and states to be polynomial in 
the size of the problem.
Each state $s$ of the gadget defines a labeled directed graph on the locations,
where a directed edge $(a,b)$ with label $s'$ means that a robot
can enter the gadget at location $a$ and exit at location~$b$, and that such a
traversal forcibly changes the state of the gadget to~$s'$.
Equivalently, a gadget is specified by its \emph{transition graph},%
\footnote{In \cite{Toggles_FUN2018}, the transition graph is called the
  ``state space'', but we feel that ``transition graph'' more clearly captures
  the automaton nature of transitions, which are discrete and directed.}
a directed graph whose vertices are state/location pairs,
where a directed edge from $(s,a)$ to $(s',b)$ represents that the robot
can traverse the gadget from $a$ to $b$ if it is in state~$s$,
and that such traversal will change the gadget's state to~$s'$.
Gadgets are \emph{local} in the sense that traversing a gadget does
not change the state of any other gadgets.

A \emph{system of gadgets} consists of gadgets, their initial states, and
a \emph{connection graph} on the gadgets' locations.
\footnote{In \cite{Toggles_FUN2018}, locations could only be matched to exactly one other location and a `branching hallway' gadget was introduced to fulfill the need of the connection graph.} If two locations $a,b$ of two gadgets (possibly the same gadget) are connected
by a path in the connection graph, then the robot can traverse freely between
$a$ and~$b$ (outside the gadgets).
(Equivalently, we can think of locations $a$ and $b$ as being identified,
effectively contracting connected components of the connection graph.)
These are all the ways that the robot can move: exterior to gadgets using
the connection graph, and traversing gadgets according to their current states.

We define a general family of \emph{motion planning} problems involving
one or more robots, each with their own start and goal location,
in a system of gadgets. In a \emph{one-player game}, we are given a system of gadgets, a single
robot that starts at a specified start location, and we want to decide whether
there is a sequence of moves that brings the robot to a specified goal location.
(This problem is perhaps the most common setting for robot motion planning.)

In a \emph{two-player game}, we are given a system of gadgets and the start
and goal locations of two robots, and two players alternate moving
their own robot by traversing a single gadget (entering at a location reachable
from the robot's current location via the connection graph).
Both players have complete information about the locations of the robots, the locations of the gadgets, and the states of the gadgets.
Here we count gadget traversals as costing one move,
and view movement in the connection graph as instantaneous/free.
The goal is to decide whether the first player has a \emph{forced win}, that is,
their robot can reach their goal location before the second player's does,
no matter how the second player responds to the first player's moves.
In a \emph{team game}, there are more than two robots, each controlled by a
single player, and the robots/players are partitioned into two teams;
the goal of each team is to get any one of its player's robot to their
goal location.  Crucially, after a team game begins, each player has only
partial information of the current gadgets' states: they can only see the state
of the gadgets reachable by their robot via the connection graph.

We also define \emph{planar motion planning}. In this case, the cyclic order of locations on a gadget is specified, and the system of gadgets must be embedded in the plane without intersections. Specifically, construct the following graph from a system of gadgets: replace each gadget with a wheel graph, which has a cycle of vertices corresponding to the locations on the gadget in the appropriate order, and a central vertex connected to each location. Connect locations on these wheels with edges according to the connection graph. The system of gadgets is \emph{planar} if this graph is planar. In planar motion planning, we restrict the problem to planar systems of gadgets. Note that this allows rotations and reflections of gadgets, but no other permutation of their locations. In some contexts, one may want to disallow reflections of gadgets, which corresponds to imposing a handedness constraint on the planar embedding of each wheel. 

\subsection{Gadget Types}

We define different subclasses of gadgets that naturally model motion planning
where the number of moves is either polynomially bounded or unbounded
(potentially exponential).

In both cases, we require that the various states of a gadget
differ only in their orientations of the possible traversals.
More precisely, a \emph{$k$-tunnel} gadget has $2k$ locations,
paired in a perfect matching whose pairs are called \emph{tunnels}, such that
each state defines which direction(s) each tunnel can be traversed. 
All of the gadgets we consider in this paper are $k$-tunnel.

In the polynomial case, we focus on ``DAG'' gadgets.
First define the \emph{state-transition (multi)graph} of a gadget to have
a vertex for each state, and a directed edge from $s$ to $s'$ for each
possible traversal of the gadget in state $s$ that leads to state~$s'$.
(This graph can be obtained from the transition graph
by combining together all vertices with the same state.)
Then a gadget is a \emph{DAG}
if its state-transition graph is a directed acyclic graph.
Such gadgets naturally lead to polynomially bounded motion planning,
as every gadget traversal consumes potential within that gadget,
as measured by the state
(e.g., in a topological ordering of the state-transition graph).
The total number of traversals is thus bounded by the total number of states
in all gadgets in the system.
(It is not enough to require that the transition graph be acyclic, because the
 robot can use the connection graph and other gadgets to reach other locations
 of this gadget in between traversals.)

In the polynomially unbounded case, we focus on gadgets that
are ``deterministic'' and ``reversible''.
A gadget is \emph{deterministic} if its transition graph has maximum out-degree
$\leq 1$; i.e., a robot entering the gadget at some location in some
state can exit at only one location and in only one state.
A gadget is \emph{reversible} if its transition graph has the reverse of every
edge, i.e., it is the bidirectional version of an undirected graph.
Thus a robot can immediately undo any gadget traversal.%
\footnote{This notion is different than the sense of ``reversible''
  in reversible computing, which would mean that we could derive
  which move to undo from the current state; here the undoing move
  only needs to be an option.}
Together, determinism and reversibility are equivalent to requiring that
the transition graph is the bidirectional version of a matching.

We also consider \emph{planar motion planning} problems with a
\emph{planar} system of gadgets, where the gadgets and
connections are drawn in the plane without crossings.  Planar gadgets are
drawn as small regions (say, disks) with their locations as points
in a fixed clockwise order along their boundary.  A single gadget type thus
corresponds to multiple planar gadget types, depending on the choice of the
clockwise order of locations.  Connections are drawn as paths connecting
the points corresponding to the endpoint locations, without crossing
gadget interiors or other connections.

The gadget model described above is an extension of the model introduced in
\cite{Toggles_FUN2018}, which characterized $2$-state deterministic reversible
$k$-tunnel gadgets that make for PSPACE-complete one-player games
(polynomially unbounded), and showed that this characterization is the same
when restricting to planar systems of gadgets.  This prior result corresponds
to the $2$-state special case of our result in the bottom-left cell of
Table~\ref{results}.  In this paper, we generalize that characterization to
gadgets with arbitrarily many states, and generalize to 2-player games,
team games, and (polynomially bounded) DAG $k$-tunnel gadgets.

\subsection{Our Characterizations}

In each type of motion planning problem where we obtain a full
characterization of easy vs.\ hard gadget sets (bounded one-player, bounded
two-player, bounded team, and unbounded one-player), we identify a class of
gadgets such that motion planning with any single gadget in that class is
hard, while motion planning with any collection of gadgets not in the class is
easy. Thus, we do not see a difference in hardness between one and multiple
gadget types; it is not possible for two ``easy'' gadgets to combine into a
hard motion planning problem. This result is in surprising contrast to
Constraint Logic where multiple gadgets were required for hardness in any
setting.

For one-player motion planning, the key property of a gadget is
\emph{interacting tunnels}: the traversal of some tunnel must affect the
traversability of some other tunnel in the same gadget.\footnote{This is roughly what \cite{Toggles_FUN2018} calls `non-trivial' gadgets.}
In the unbounded case (Section~\ref{sec:1-Player Unbounded}),
we show that any such gadget (that is also reversible and deterministic) can
be used to simulate two specific gadgets, the ``locking 2-toggle'' and
``1-toggle'' (shown in Figure~\ref{fig:basic gadgets}),
which together suffice to prove PSPACE-hardness.
This argument involves surprisingly little case analysis, in contrast to the
prior work in this area \cite{Toggles_FUN2018}, which simply enumerated and
analyzed all $2$-state gadgets. On the other hand, we show that any fixed
collection of gadgets without interacting tunnels reduces (via a shortcutting
argument) to graph traversal, which can be solved in NL.
We furthermore show that this dichotomy still holds for 1-player planar motion planning (Section~\ref{sec:1-Player Unbounded Planar}). 
In the bounded case (Section~\ref{sec:1-Player Bounded}),
we examine the naturally bounded class of DAG gadgets.
We again obtain a somewhat more complicated full characterization,
which mostly depends on the existence of interacting tunnels.

For two-player motion planning, it turns out that interacting tunnels
are not required for hardness. In the bounded case
(Section~\ref{sec:2-Player Bounded}), we show that PSPACE-completeness
holds for any DAG gadget that is \emph{nontrivial}, i.e., has at least
one transition in some state. We show that any nontrivial DAG gadget
can simulate one of two one-tunnel gadgets, ``single-use
unidirectional edge'' or ``single-use bidirectional edge'', and
surprisingly either suffices to prove PSPACE-completeness. A single
use-use edge is a transition in a gadget such that after taking that
transition, there are no further transitions between the two
associated locations. Obviously, two-player motion planning with
trivial gadgets is in P: the robots can only traverse the connection
graph, and one merely needs to see which is closer to their goal.  In
the unbounded case (Section~\ref{sec:2-Player Unbounded}), we show
that any gadget with interacting tunnels suffices for
EXPTIME-completeness, and it remains an open problem whether some
weaker condition suffices.

For team motion planning, interacting tunnels are again not required for
hardness. In the bounded case (Section~\ref{sec:Team Bounded}),
we show that NEXPTIME-completeness holds for
any nontrivial DAG gadget, again by showing that any single-use edge gadget
suffices. In the unbounded case (Section~\ref{sec:Team Unbounded}),
we again show that any gadget with
interacting tunnels suffices for undecidability, and it remains an open
problem whether some weaker condition suffices.

Armed with the general framework of this paper, it should be much easier to
prove hardness of most games that involve motion planning of robots in an
environment with nontrivial local state.  You simply need to pick a gadget
that is hard according to our characterization (with the matching boundedness
and number of players/teams), draw a single figure of how to build that gadget
within the game of interest, and check that it is possible to connect these
gadgets together.  While this paper focuses on general theory building, we
return to possible applications in Section~\ref{sec:applications}.

\section{1-Player Unbounded Motion Planning}
\label{sec:1-Player Unbounded}
In this section, we study reversible, deterministic gadgets, extending the work in \cite{Toggles_FUN2018} which only considered gadgets with two or fewer states. Here we give a complete categorization as either in NL or PSPACE-complete for reversible, deterministic gadget. For the NL half of the characterization, Theorem~\ref{thm:trivial} below shows that 1-player motion planning problems with non-interacting-$k$-tunnel gadgets is in NL. For the PSPACE-completeness half of the characterization, we introduce a new base gadget, the \emph{locking 2-toggle (L2T)} shown in Figure~\ref{fig:L2T}. In Section~\ref{sec:reducing to locking 2-toggles} we show that all interacting-$k$-tunnel reversible deterministic gadgets are able to simulate the locking 2-toggle. Then in Section~\ref{sec:NCL to 2-toggle} we show that 1-player motion planning with locking 2-toggles is PSPACE-complete by simulating Nondeterministic Constraint Logic. Section~\ref{sec:NCL to 2-toggle} shows how to adapt the construction to show these gadgets remain PSPACE-hard even for the planar 1-player motion planning problem.

\begin{lemma}\label{lem:all gadgets PSPACE}
  1-player motion planning with any set of gadgets
  is in PSPACE.
\end{lemma}
\begin{proof}
  This was shown in \cite{Toggles_FUN2018}, but included here for convenience. A configuration of the system of gadgets consists of the state of each gadget and the location of the robot, and has polynomial length. The algorithm that repeatedly nondeterministically picks a legal transition, and updates the configuration based on it, accepting when the robot reaches the goal location, decides the reachability problem in nondeterministic polynomial space. By Savitch's theorem, the problem is in PSPACE.
\end{proof}

\begin{theorem}\label{thm:trivial}
  1-player motion planning with any $k$-tunnel gadget that does not have interacting tunnels is in NL.
\end{theorem}

\begin{proof}
  We first show that if a system of such gadgets has a solution, then it has a solution which visits each location at most once. Suppose there is a solution, and consider the last time a solution of minimal length visits a previously visited location, assuming there is any such time. Let $v$ be the vertex of this last self-intersection. After leaving $v$ for the last time, every transition the robot makes is through a tunnel that it had not previously traversed. Since the gadget does not have interacting tunnels, these tunnels have the same traversability when the robot goes through them as they do originally. We modify the solution by `shortcutting': remove the portion of the solution between the first visit to $v$ and the last visit to $v$, so the robot only visits $v$ once, and skips the loop that begins and ends at $v$. The new path is still a solution: the segment before $v$ is identical to the unmodified solution, and the segment after $v$ consists of tunnels whose traversability is not changed before the robot goes through them. The shortcut path is shorter than the original solution, which was assumed to be minimal. Thus a solution of minimal length has no self-intersections.

  We'll want to treat the system of gadgets as though it were a directed graph by replacing each tunnel with an edge in the appropriate direction, or a pair edges if it is traversable in either direction. We can locally walk through all the available transitions in a gadget, assess which locations they lead to, and non-deterministically pick one to try, allowing this to be executed in NL. A path from the start location to the end location in this graph is exactly a solution for the system of gadgets with no self-intersections; the traversability of each tunnel used in such a solution does not change before the tunnel is used. 

  Since reachability in directed graphs is in NL, the motion planning problem is also in NL. Moreover, if the gadget has any state in which a tunnel can be traversed in one direction but not the other, the motion planning problem is NL-complete, and otherwise it is in L.
\end{proof}

\subsection{Reducing to Locking 2-Toggles}
\label{sec:reducing to locking 2-toggles}

In this section, we introduce the locking 2-toggle shown in Figure~\ref{fig:L2T}, and we show that all interacting-$k$-tunnel reversible deterministic gadgets can simulate it. The proof first examines what constraints on a gadget are implied by being interacting-$k$-tunnel, reversible, and deterministic, and goes on to identify that all such gadgets have a pair of special states with some useful common properties. From this pair of states we construct a 1-toggle, and then combine that with our special states to build a locking 2-toggle. One of the major insights is identifying this special pair of states which belongs to all gadgets in the class, and after that the primary challenge is in preventing undesired transitions, which are plentiful when allowing such a wide class of gadgets.

\begin{theorem}\label{thm:arb gadget sim L2T}
  Every interacting-$k$-tunnel reversible deterministic gadget simulates a locking 2-toggle.
\end{theorem}

\begin{proof}
  We begin by examining an arbitrary interacting-$k$-tunnel reversible deterministic gadget, as shown in Figure~\ref{fig:arb gadget}. Because the gadget has interacting tunnels, we can find a pair of states in which traversing the top line can change the traversability of the bottom line to the right. Since it is also reversible, the inverse transition is also possible, so traversing the top line can change in either direction the left-to-right traversability of the bottom line. Then without loss of generality, the gadget has the form shown in Figure~\ref{fig:arb gadget}: in state 1, traversing the top line to the right switches to state 2, and the bottom line is not traversable to the right. In state 2, traversing the top line to the left switches to state 1, and the bottom line is traversable to the right, say to state 3 (which may be the same as state 1). All other traversals may or may not be possible in either state, indicated by the question marks. 
\begin{figure}
  \centering
  \includegraphics[scale=0.75]{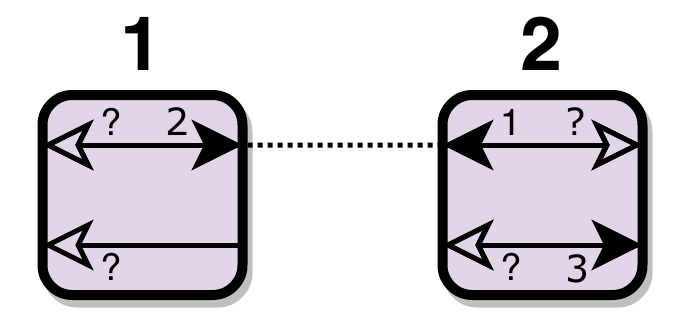}
  \caption{An arbitrary interacting-$k$-tunnel reversible deterministic gadget.
    Hollow arrows indicate traversals that may or may not be possible.
    Solid or absent arrows indicate traversals that are or are not possible,
    respectively.}
  \label{fig:arb gadget}
\end{figure}

  \begin{lemma}\label{lem:arb gadget sim 1-way}
    Every interacting-$k$-tunnel reversible deterministic gadget simulates a \emph{one-directional edge}, that is, a tunnel which (in some state) can be traversed in one direction but not the other.
  \end{lemma}

  \begin{proof}
    If in some state, some edge in the gadget can be traversed in one direction but not the other, then it is a one-directional edge. Otherwise, the gadget has the form shown in Figure~\ref{fig:arb gadget with no 1-way}. Then the construction in Figure~\ref{fig:arb gadget sim 1-way} is equivalent to a one-directional edge: currently the gadget is in state 1, so the path from the bottom to the top is blocked by the bottom edge, but from the top, you can go across the top edge, switching the gadget to state 2, and then back across the bottom edge.
  \end{proof}

\begin{figure}
  \centering
  \begin{subfigure}{0.45\textwidth}
    \centering
    \includegraphics[scale=0.75]{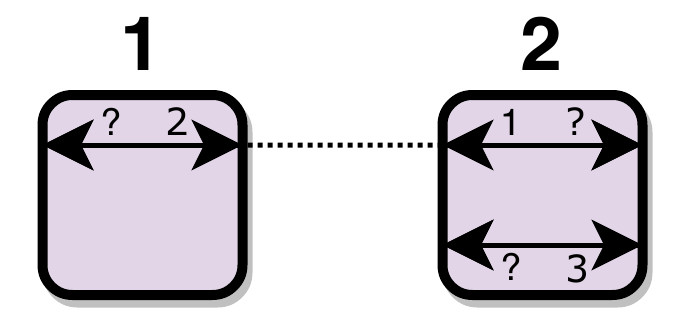}
    \caption{State graph, refining Figure~\ref{fig:arb gadget}.}
    \label{fig:arb gadget with no 1-way}
  \end{subfigure}\hfil\hfil
  \begin{subfigure}{0.45\textwidth}
    \centering
    \includegraphics[scale=0.75]{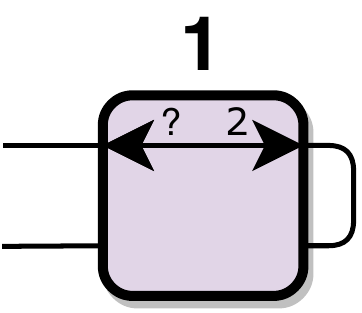}
    \caption{Simulating a one-directional edge.}
    \label{fig:arb gadget sim 1-way}
  \end{subfigure}
  \caption{An arbitrary interacting-$k$-tunnel reversible deterministic gadget which has no one-directional edge.}
\end{figure}

  \begin{lemma}\label{lem:constructing 1-toggle}
    Every interacting-$k$-tunnel reversible deterministic gadget simulates a 1-toggle (Figure~\ref{fig:1-toggle}).
  \end{lemma}

  \begin{proof}
    By the previous lemma, we can build a one-directional edge, which has the structure shown in Figure~\ref{fig:1-way edge}: in state 1, we can traverse the edge to the right and switch to state 2, but not to the left. In state 2, we can undo this transition, and possibly also traverse the edge to the right. The construction in Figure~\ref{fig:1-way sim 1-toggle} is then a 1-toggle. In the current state, it can be traversed to the right but not to the left because of the gadget on the left. After making this traversal, it becomes the rotation of the current state, and it cannot be traversed to the right again because of the gadget on the right.
  \end{proof}

\begin{figure}
  \begin{subfigure}{0.45\textwidth}
    \centering
    \includegraphics[scale=0.75]{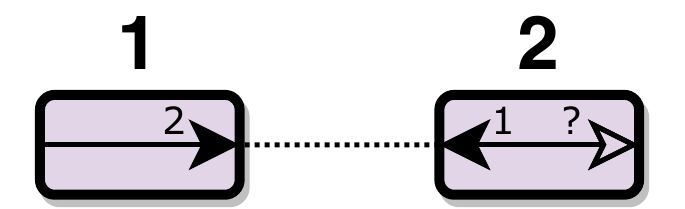}
    \caption{Form of state graph.}
    \label{fig:1-way edge}
  \end{subfigure}\hfil\hfil
  \begin{subfigure}{0.45\textwidth}
    \centering
    \includegraphics[scale=.75]{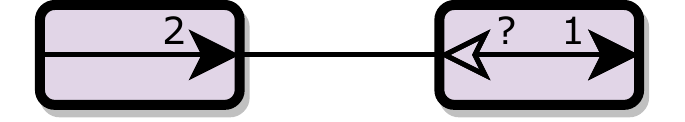}
    \caption{Simulating a 1-toggle.}
    \label{fig:1-way sim 1-toggle}
  \end{subfigure}
  \caption{A one-directional edge gadget.}
\end{figure}

  To build a locking 2-toggle, we put the arbitrary gadget (in state 2), an antiparallel pair of 1-toggles, and the rotation of the arbitrary gadget (also in state 2) in series, as in Figure~\ref{fig:arb gadget sim L2T}. Currently, the top edge is traversable to the left and the bottom edge is traversable to the right, but not in the other direction. After traversing the top edge to the left, the 1-toggles prevents us from traversing either edge to the left, and the leftmost gadget (in state 1) prevents us from traversing the bottom edge to the right, so the only legal traversal is going back across the top edge to the right. Similarly after traversing the bottom edge, the only legal traversal is across the bottom edge in the opposite direction. Thus this construction is equivalent to a (antiparallel) locking 2-toggle.

\begin{figure}
  \centering
  \includegraphics[scale=.75]{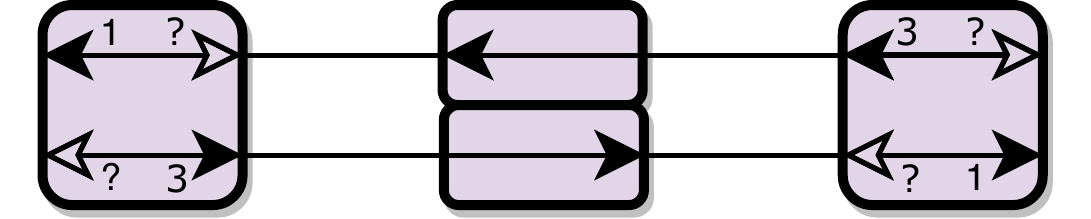}
  \caption{An arbitrary interacting-$k$-tunnel reversible deterministic gadget and a 1-toggle simulate a locking 2-toggle.}
  \label{fig:arb gadget sim L2T}
\end{figure}

  Traversing the simulated locking 2-toggle takes either 4 or 6 transitions of the raw gadget, depending on whether it contains a one-directional edge (from Lemma~\ref{lem:arb gadget sim 1-way}). For simplicity, we can include additional gadgets (e.g. another pair of 1-toggles) to ensure it always takes exactly 6 transitions; this will be relevant to timing considerations in multiplayer games.
\end{proof}

\begin{figure}
  \centering
  \begin{subfigure}{0.45\textwidth}
      \centering
      \includegraphics[scale=.47]{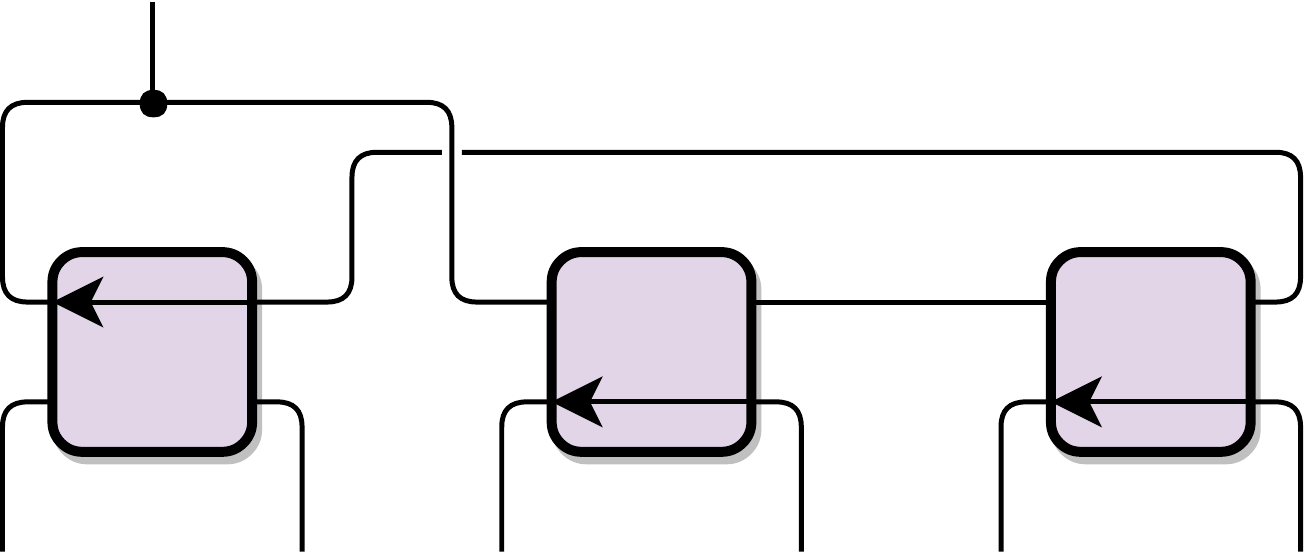}
      \caption{An AND vertex gadget. The leftmost edge has weight two and is pointing in (up). The other edges have weight one and are pointing away (down).}
      \label{fig:NCL AND vertex}
  \end{subfigure}\hfill
  \begin{subfigure}{0.45\textwidth}
      \centering
      \includegraphics[scale=.47]{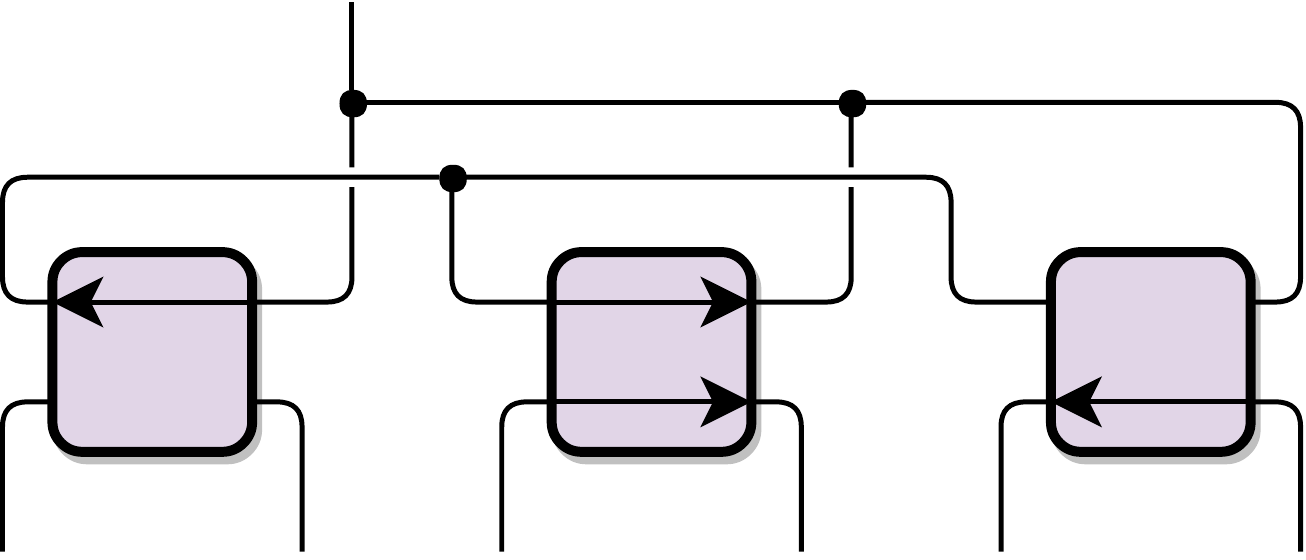}
      \caption{An OR vertex gadget. All edges are weight 2. The leftmost edge is pointing in (up), the middle edge is free, and the rightmost edge is pointing away (down).}
      \label{fig:NCL OR vertex}
  \end{subfigure}
  \caption{Vertex gadgets in the NCL reduction.}
  \label{fig:NCL vertex gadgets}
\end{figure}

\begin{figure}
  \centering
  \begin{subfigure}{0.45\textwidth}
      \centering
      \includegraphics[scale=.5]{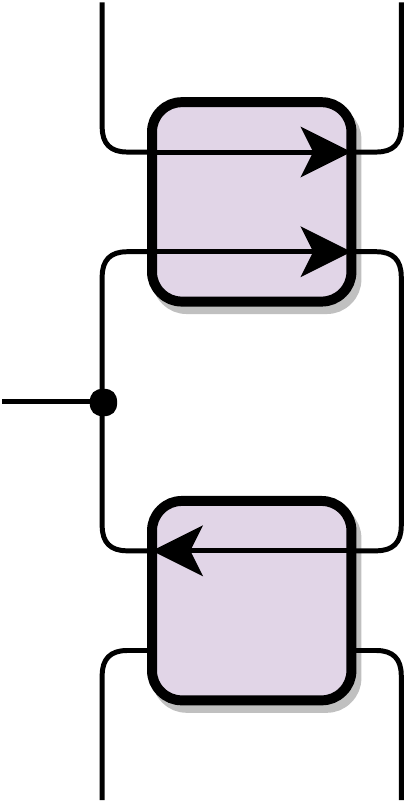}
      \caption{An edge gadget pointed up, in the unlocked state. The gadget is accessed by the loose end on the left.}
      \label{fig:NCL edge unlocked}
  \end{subfigure}\hfill
  \begin{subfigure}{0.45\textwidth}
      \centering
      \includegraphics[scale=.5]{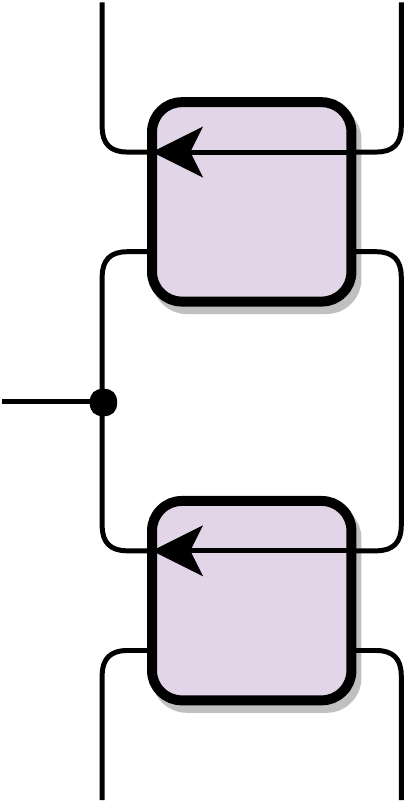}
      \caption{The same edge gadget in the locked state.}
      \label{fig:NCL edge locked}
  \end{subfigure}
  \caption{Edge gadget in the NCL reduction.}
  \label{fig:NCL edge gadget}
\end{figure}

\subsection{PSPACE-hardness}
\label{sec:NCL to 2-toggle}

In this section, we show that 1-player motion planning with the
locking 2-toggle is PSPACE-complete by a reduction from
Nondeterministic Constraint Logic (NCL). See
Appendix~\ref{app:constraint logic} for a definition of NCL. We
represent edges by pairs of locking 2-toggles. The construction requires \emph{edge gadgets} which are directed and can be flipped, as well as AND and OR \emph{vertex gadgets} which apply constraints on how many edges must be directed towards them at any given point in time.

\begin{theorem}\label{thm:L2T PSPACE-c}
  1-player motion planning with the locking 2-toggle is PSPACE-complete.
\end{theorem}

\begin{proof}
  Motion planning with the gadget is in PSPACE by Lemma~\ref{lem:all gadgets PSPACE}. We use a reduction from Nondeterministic Constraint Logic (NCL) to show PSPACE-hardness.
  See Appendix~\ref{app:constraint logic} for a definition of NCL.

The \textbf{edge gadget}, shown in Figure~\ref{fig:NCL edge gadget}, contains two locking 2-toggles, each of which is also attached to a vertex gadget. It is oriented towards one of the vertices, can be either \emph{locked} or \emph{unlocked}. Specifically, the edge gadget is unlocked (Figure~\ref{fig:NCL edge unlocked}) if either locking 2-toggle is in the middle state (with both lines traversable), and locked (Figure~\ref{fig:NCL edge locked}) otherwise. It is oriented towards the vertex attached to the locking 2-toggle whose edge not accessible from the edge gadget is traversable. The robot can access the free line on the left. If the edge gadget is unlocked, the robot can traverse a loop through one edge of each locking 2-toggle to change the orientation of the edge gadget. The edge gadget switches between being locked and unlocked when the robot moves through a vertex gadget to traverse one of the edges not accessible from the edge gadget.

  The \textbf{vertex gadgets} are shown in Figure~\ref{fig:NCL vertex gadgets}. The robot can access the free line on the top, and traverse loops to lock and unlock edge gadgets, enforcing the constraints of vertices. Specifically, if all three edges are pointing towards an AND vertex, the robot can traverse a loop to lock both weight-1 edges and unlock the weight-2 edge, or vice versa. If multiple edges are pointing towards an OR vertex, the robot can traverse a loop to unlock the currently locked edge and lock another edge. Observe that for both vertex gadgets, the sum of the weights of locked edges does not change.

  Given an NCL graph, we construct a maze of locking 2-toggles. Each edge in the graph corresponds to an edge gadget (Figure~\ref{fig:NCL edge gadget}). Each locking 2-toggle in the edge gadget corresponds to a vertex incident to the edge. When three edges meet at a vertex, we put a vertex gadget on the locking 2-toggles corresponding to that vertex. We use an AND vertex gadget (Figure~\ref{fig:NCL AND vertex}) or an OR vertex gadget (Figure~\ref{fig:NCL OR vertex}) depending on the type of vertex. The vertical `entrance' line on each vertex gadget and horizontal `entrance' line on each edge gadget is connected to the starting location. Each edge is oriented as in the NCL graph. For each vertex, we pick a set of edges initially pointing at the vertex with total weight 2. The edge gadgets corresponding to the chosen edges are locked, and other edge gadgets are unlocked. The goal location is placed inside the edge gadget corresponding to the target edge so that it is reachable if and only if the target edge is unlocked.

  If the original NCL graph is solvable, the robot can perform the same sequence of edge flips, visiting vertex gadgets to lock and unlock edges as necessary, and reach the goal location. If the robot can reach the goal location, the same sequence of edge flips solves the NCL graph. So the maze is solvable if and only if the NCL graph was.
\end{proof}

This reduction is also possible without edge gadgets, and leads to a system with only one L2T for each constraint logic edge. We use edge gadgets because the reduction is easier to understand, and adaptations of this construction in Sections~\ref{sec:1-Player Unbounded Planar}, \ref{sec:2-Player Unbounded}, and \ref{sec:Team Unbounded} will need them.

\begin{corollary}\label{cor:arb gadget PSPACE-c}
  1-player motion planning with
  any interacting-$k$-tunnel reversible deterministic gadget is PSPACE-complete.
\end{corollary}

\begin{proof}
  Hardness follows from Theorems~\ref{thm:arb gadget sim L2T}, and \ref{thm:L2T PSPACE-c}. For any such gadget, we have a reduction from mazes of locking 2-toggles to mazes of that gadget by replacing each locking 2-toggle with a simulation of one built from the arbitrary gadget. Motion planning with the gadget is in PSPACE by Lemma~\ref{lem:all gadgets PSPACE}.
\end{proof}

\subsection{Planarity}
\label{sec:1-Player Unbounded Planar}
In this section, we show that interacting-$k$-tunnel reversible deterministic gadgets are PSPACE-complete even for the planar 1-player motion planning problem. We once again work with the locking 2-toggle, showing that each of its planar versions can simulate each other. From there we use the crossing locking 2-toggle to build an A / BA crossover, which is less powerful than a full crossover but will suffice to make our reduction in Section~\ref{sec:NCL to 2-toggle} planar. An interesting question is whether the locking 2-toggle is powerful enough to build a full crossover, which can be done with any of the 2 state gadgets. Although not needed here, it would allow the multiplayer game results later in this paper to carry over to the planar case.

\begin{figure}
  \centering
  \begin{subfigure}{.3\textwidth}
      \centering
      \includegraphics[scale=.65]{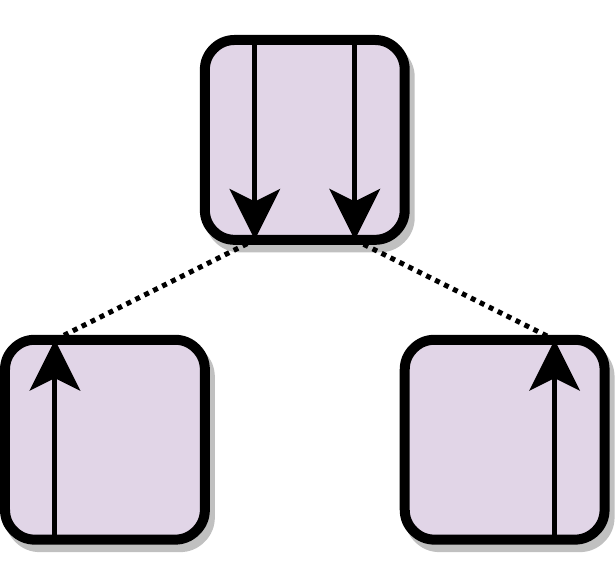}
      \caption{A parallel locking 2-toggle (PL2T).}
      \label{fig:PL2T}
  \end{subfigure}
  \begin{subfigure}{.3\textwidth}
      \centering
      \includegraphics[scale=.65]{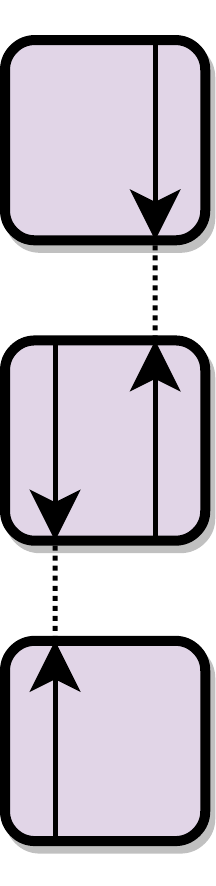}
      \caption{An antiparallel locking 2-toggle (APL2T).}
      \label{fig:APL2T}
  \end{subfigure}
  \begin{subfigure}{.3\textwidth}
      \centering
      \includegraphics[scale=.65]{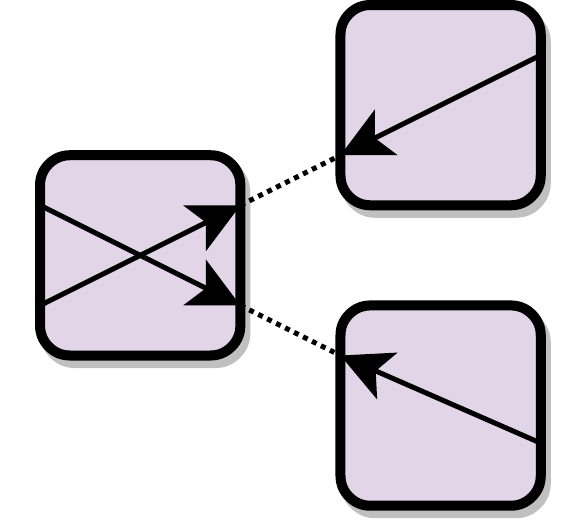}
      \caption{A crossing locking 2-toggle (CL2T).}
      \label{fig:CL2T}
  \end{subfigure}
  \caption{Types of locking 2-toggles in planar mazes.}
  \label{fig:L2T types}
\end{figure}

Recall for the planar problem we allow rotations and reflections of gadgets. This leaves three distinct embeddings of the locking 2-toggle into a plane: parallel, antiparallel, and crossing, shown in Figure~\ref{fig:L2T types}, and which we abbreviate PL2T, APL2T, and CL2T. (Up to only rotation, there are four, the other being the antiparallel locking 2-toggle with the other handedness). We will allow reflections of gadgets, so these are the three kinds of locking 2-toggles we will consider.

\begin{lemma}[\cite{Toggles_FUN2018}] \label{lem:L2Ts equivalent}
  Parallel, antiparallel, and crossing locking 2-toggles all simulate each other in planar graphs.
\end{lemma}

\begin{figure}
  \centering
  \begin{minipage}{0.43\linewidth}
    \centering
    \includegraphics[scale=0.65]{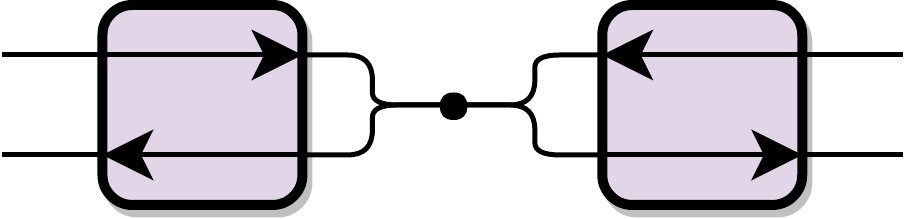}
    \caption{APL2T simulating CL2T. (Based on \cite[Figure~4]{Toggles_FUN2018}.)}
    \label{fig:APL2T sim CL2T}
  \end{minipage}\hfil\hfil
  \begin{minipage}{0.43\linewidth}
    \centering
    \includegraphics[scale=0.65]{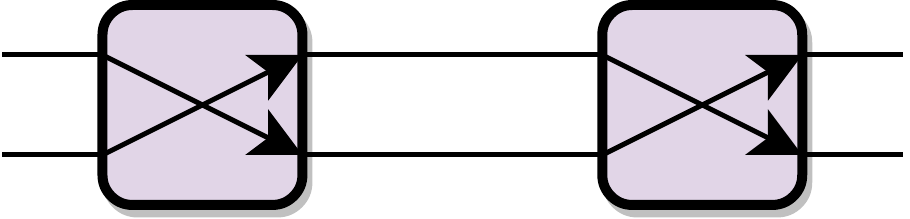}
    \caption{CL2T simulating PL2T. (Based on \cite[Figure~5]{Toggles_FUN2018}.)}
    \label{fig:CL2T sim PL2T}
  \end{minipage}
\end{figure}

\begin{figure}
  \centering
  \includegraphics[scale=0.75]{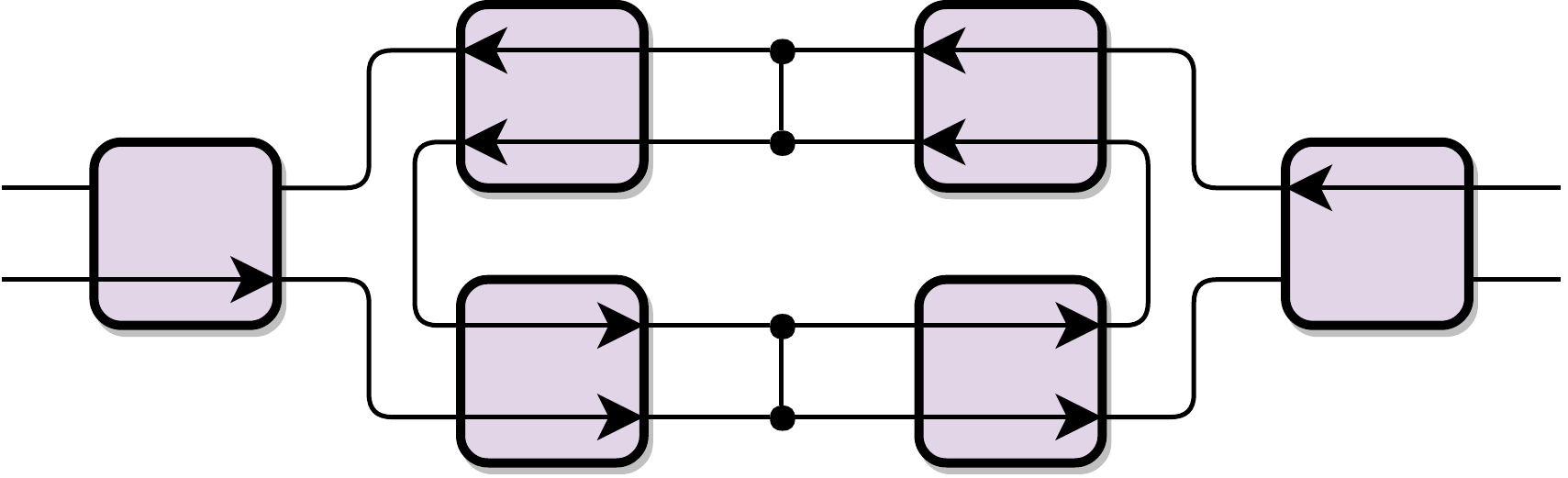}
  \caption{PL2T simulating APL2T. (Based on \cite[Figure~13]{Toggles_FUN2018}.)}
  \label{fig:PL2T sim APL2T}
\end{figure}

\begin{proof}
  Figure~\ref{fig:APL2T sim CL2T} shows APL2T simulating CL2T, Figure~\ref{fig:CL2T sim PL2T} shows CL2T simulating PL2T, and Figure~\ref{fig:PL2T sim APL2T} shows PL2T simulating APL2T. Note that we use both APL2Ts of both handednesses, so we need to be able to reflect gadgets.
\end{proof}


\begin{theorem}\label{thm:planar arb sim L2T}
  Every interacting-$k$-tunnel reversible deterministic gadget simulates each type of locking 2-toggle in planar graphs.
\end{theorem}

\begin{figure}
  \centering
  \includegraphics[scale=.75]{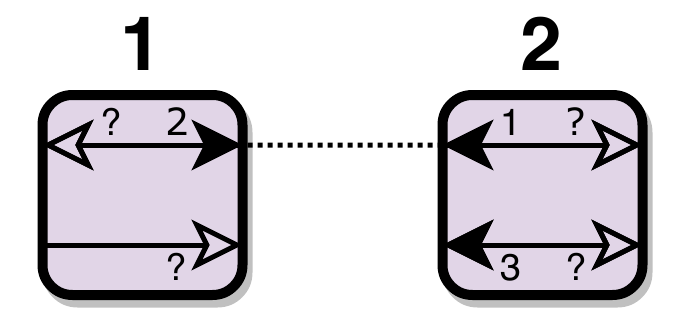}
  \caption{The antiparallel case of an arbitrary interacting-$k$-tunnel reversible deterministic gadget.}
  \label{fig:P arb gadget}
\end{figure}

\begin{figure}
  \centering
  \includegraphics[scale=.75]{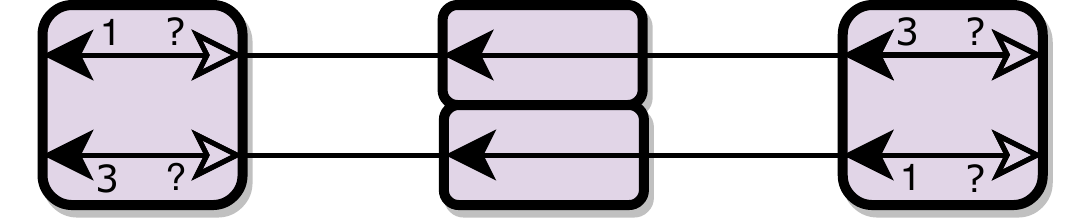}
  \caption{An arbitrary antiparallel interacting-$k$-tunnel reversible deterministic gadget and a 1-toggle simulate a PL2T.}
  \label{fig:P arb gadget sim PL2T}
\end{figure}

\begin{figure}
  \centering
  \includegraphics[scale=.75]{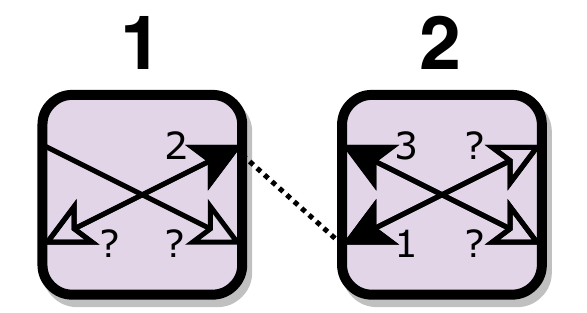}
  \caption{The crossing case of an arbitrary interacting-$k$-tunnel reversible deterministic gadget.}
  \label{fig:C arb gadget}
\end{figure}

\begin{figure}
  \centering
  \includegraphics[scale=.75]{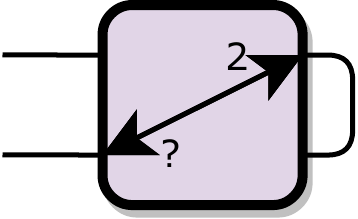}
  \caption{A crossing interacting-$k$-tunnel reversible deterministic gadget simulates a one-way edge.}
  \label{fig:C arb gadget sim 1-way}
\end{figure}

\begin{figure}
  \centering
  \includegraphics[scale=.75]{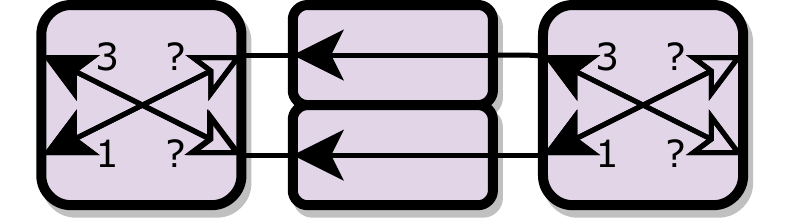}
  \caption{An arbitrary crossing interacting-$k$-tunnel reversible deterministic gadget and a one-toggle simulate a PL2T.}
  \label{fig:C arb gadget sim L2T}
\end{figure}

\begin{proof}
  We follow the proof of Theorem~\ref{thm:arb gadget sim L2T}. As before, we assume that traversing a line to switch from state 1 to state 2 makes a traversal on another line legal. This new traversal can be parallel to, antiparallel to, or cross the first traversal; we consider each case. If the new traversal is parallel, the construction in the proof of Theorem~\ref{thm:arb gadget sim L2T} works to simulate an APL2T in a planar graph. 

  If it is antiparallel, the gadget has the form shown in Figure~\ref{fig:P arb gadget}. Either the gadget has a one-directional edge, or it has the form in Figure~\ref{fig:arb gadget with no 1-way}, and simulates a one-directional edge by the construction in Figure~\ref{fig:arb gadget sim 1-way}. Thus it simulates a 1-toggle by the construction in Figure~\ref{fig:1-way sim 1-toggle}. Then the construction in Figure~\ref{fig:P arb gadget sim PL2T} simulates a PL2T: currently either edge can be traversed to the left, if the top edge is traversed, the left gadget blocks the bottom edge, and if the bottom edge is traversed, the right gadget blocks the top edge.

  Finally, if the new traversal crosses the first traversal, the gadget has the form shown in Figure~\ref{fig:C arb gadget}. Either it has a one-directional edge, or the construction in Figure~\ref{fig:C arb gadget sim 1-way} simulates a one-directional edge, similarly to Lemma~\ref{lem:arb gadget sim 1-way}. So the gadget simulates a 1-toggle by the construction in Figure~\ref{fig:1-way sim 1-toggle}. Then the construction in Figure~\ref{fig:C arb gadget sim L2T} simulates a PL2T, similarly to the previous case.

  Once the gadget simulates some locking 2-toggle, we can use Lemma~\ref{lem:L2Ts equivalent} to simulate all three types.
\end{proof}

\begin{theorem}\label{thm:planar 1p unbounded}
  1-player planar motion planning with
  any interacting-$k$-tunnel reversible deterministic gadget
  is PSPACE-complete.
\end{theorem}

\begin{figure}
  \centering
  \begin{subfigure}{.45\textwidth}
    \centering
    \includegraphics[scale=.75]{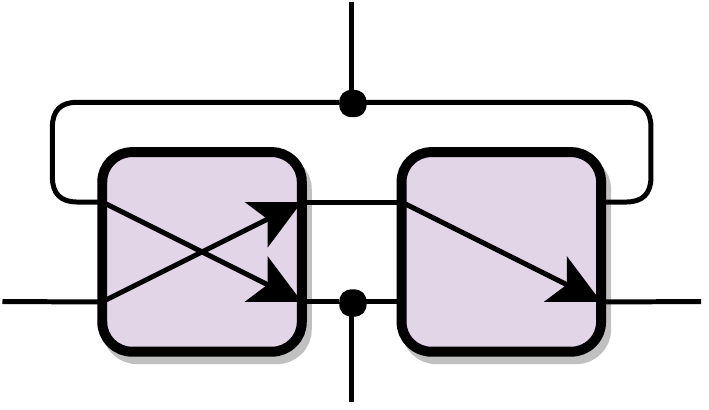}
    \caption{Simulating an A/BA crossover using CL2Ts.}
    \label{fig:CL2T sim A/BA}
  \end{subfigure}
  \begin{subfigure}{.45\textwidth}
    \centering
    \includegraphics[scale=.75]{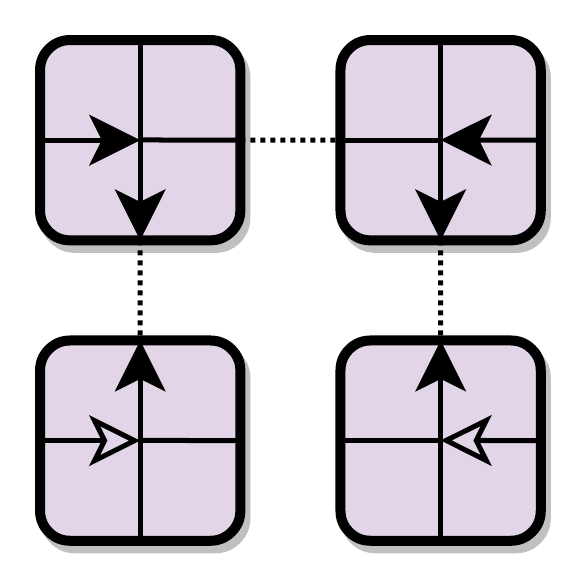}
    \caption{A state diagram and notation for the A/BA crossover.}
    \label{fig:A/BA notation}
  \end{subfigure}
  \caption{An A/BA crossover gadget: the robot can traverse top to bottom (A), or traverse left to right (B) and then top to bottom. Thinking of the gadget as a crossing pair of 1-toggles, the vertical 1-toggle is always traversable, and the horizontal 1-toggle is traversable when the vertical one is pointing down.}
  \label{fig:A/BA crossover}
\end{figure}

\begin{figure}
  \centering
  \includegraphics[scale=.75]{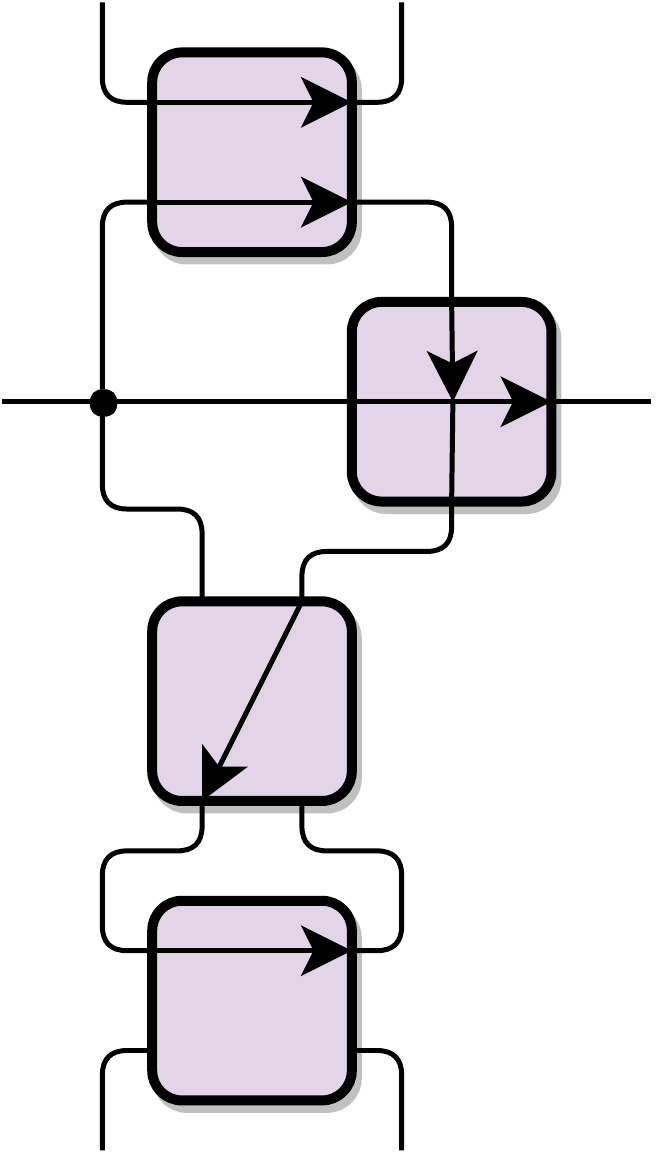}
  \caption{An edge gadget for planar graphs, currently unlocked and directed up. This is analogous to Figure~\ref{fig:NCL edge gadget}, with two changes. First, the bottom PL2T is `twisted' to have the same handedness as the top PL2T for connecting to vertex gadgets; the CL2T is sufficient for the crossing caused by this. Second, the A/BA crossover allows the robot to cross the edge from left to right, regardless of the state of the edge. We call the line on the left the \emph{entrance} and the line on the right, on the other side of the A/BA crossover, the \emph{exit}.}
  \label{fig:planar NCL edge gadget}
\end{figure}

\begin{figure}
  \centering
  \begin{subfigure}{.4\textwidth}
    \centering
    \includegraphics[scale=.74]{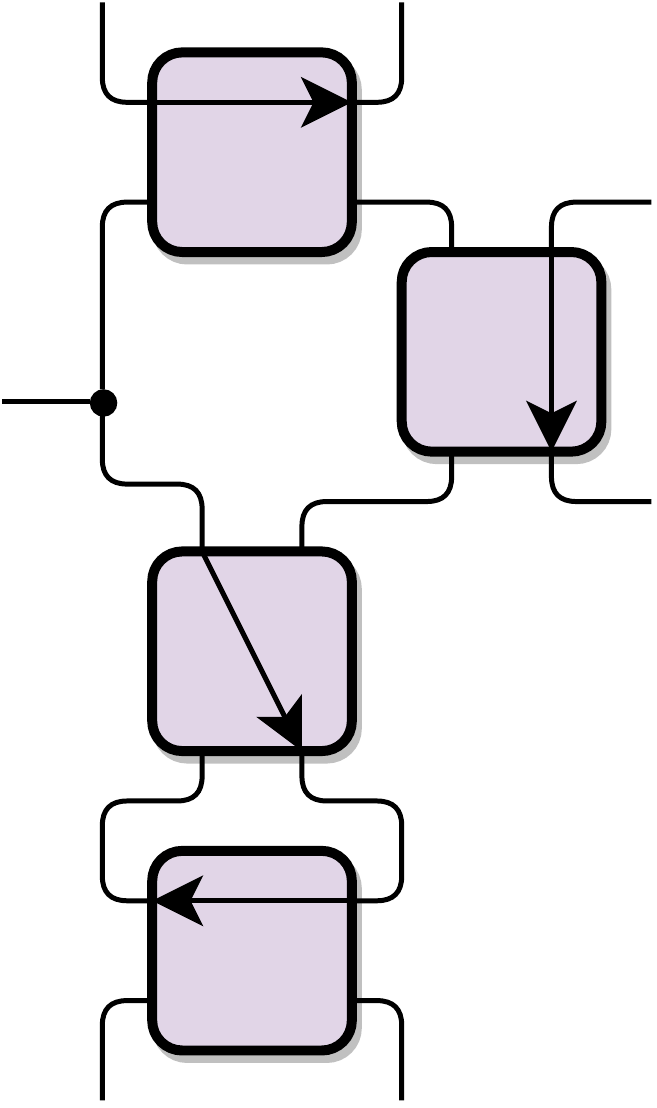}
    \caption{An AND vertex for planar graphs. Currently the weight-2 edge, connected at the bottom PL2T, is directed towards the vertex and locked, and both weight-1 edges are directed away. If the weight-1 edges become directed towards the vertex, the robot can visit the vertex gadget and traverse a loop through all three PL2Ts, locking the weight-1 edges and unlocking the weight-2 edge. The CL2T is a sufficient crossover.}
    \label{fig:planar NCL AND vertex}
  \end{subfigure}\hfil\hfil
  \begin{subfigure}{.5\textwidth}
    \centering
    \includegraphics[scale=.75]{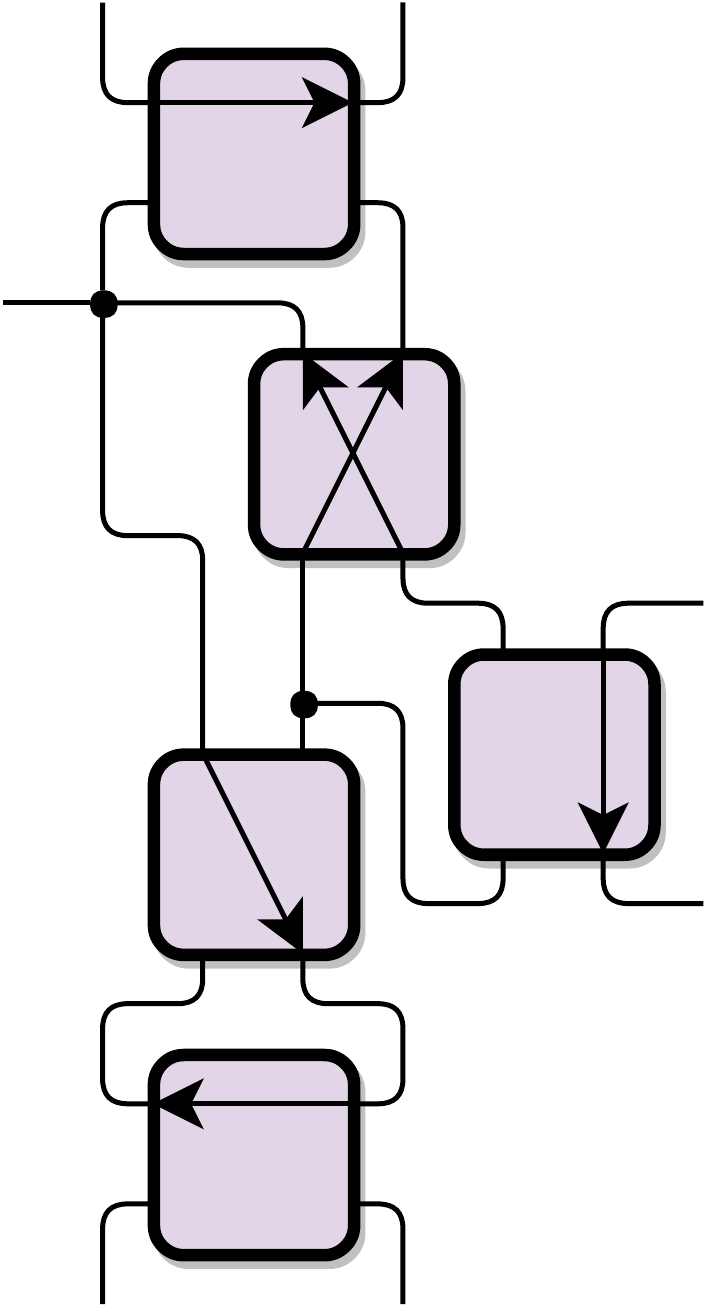}
    \caption{An OR vertex for planar graphs. Currently the edge containing the bottom PL2T is directed towards the vertex, and the other edges are directed away. If multiple edges are ever directed towards the vertex, the robot can visit the vertex gadget, unlock the locked edge, and lock another edge.}
    \label{fig:planar NCL OR vertex}
  \end{subfigure}
  \caption{NCL vertex gadgets for planar graphs, analogous to the gadgets in Figure~\ref{fig:NCL vertex gadgets}. In each gadget, each of the three PL2Ts is also part of an edge gadget. The robot enters at the line on the left, called the \emph{entrance}, traverses loops that enforce the NCL constraints, and then leaves at the entrance.}
  \label{fig:planar NCL vertex gadgets}
\end{figure}

\begin{proof}
  We begin by constructing some weak crossover gadgets. The crossover locking 2-toggle is itself a very weak crossover. We use it to construct an \emph{A/BA crossover}, shown in Figure~\ref{fig:CL2T sim A/BA}. Calling the traversal from top to bottom A and that from left to right B, we can perform either of the sequences A and BA. Since everything is reversible and deterministic, we can also undo those sequences. The A/BA crossover is sufficient for the rest of the proof; we abbreviate it as shown in Figure~\ref{fig:A/BA notation}.

  We modify the proof of Theorem~\ref{thm:L2T PSPACE-c}, giving a reduction from planar NCL to planar mazes with locking 2-toggles. By Theorem~\ref{thm:planar arb sim L2T}, this is sufficient to show PSPACE-hardness. Our gadgets use PL2Ts, CL2Ts, and A/BA crossovers; they do not use APL2Ts.

  The edge gadget is shown in Figure~\ref{fig:planar NCL edge gadget}, and vertex gadgets are shown in Figure~\ref{fig:planar NCL vertex gadgets}. Given a planar NCL graph, we construct a mazes as follows.

  Pick a rooted spanning tree of the dual of the NCL graph, directed away from the root; the robot will use this tree to navigate the graph. The system of gadgets will contain a vertex for each face $f$ of the NCL graph, which is a vertex of the spanning tree.

  For each edge of the graph, we place an edge gadget. When an edge is in the spanning tree, we orient it so that the A/BA crossover points, from entrance to exit, in the same direction as the edge points in the spanning tree (left to right in Figure~\ref{fig:planar NCL edge gadget}, and away from the root). If an edge is in the spanning tree and has target $f$, we connect its exit to $f$. For each edge $e$, we connect its entrance to the vertex $f$ corresponding to the face containing its entrance, i.e. the face adjacent to $e$ to which we can connect its entrance without crossings. If $e$ is in the spanning tree, this connects the entrance of $e$ to the source $f$ of $e$.

  Now we place a vertex gadget of the appropriate type for each vertex of the NCL graph, so that the gadget shares a PL2T with each incident edge gadget. AND vertex gadgets must be oriented so the weight-2 edge has the appropriate PL2T (the bottom one in Figure~\ref{fig:planar NCL AND vertex}). The entrance of each vertex gadget is connected to the vertex $f$ corresponding to the face containing the entrance.

  We set each edge gadget to the orientation of its corresponding edge. For each vertex, we select edges directed towards it with total weight 2, and set the selected edges to locked and other edges to unlocked. The goal location is placed inside the target edge so that reaching it requires flipping the target edge. The starting location is the vertex corresponding to the root of the spanning tree.

  Play on this maze proceeds as follows: the robot travels down the spanning tree, crossing edges until it reaches some face. It goes into an edge or vertex attached to that face, and manipulates it. Then the robot travels back up the spanning tree and down a different branch, manipulating another edge or vertex, and so on. The edge and vertex gadgets enforce the NCL constraints. If the target edge can be flipped, the robot can reach the goal location. Thus the maze is solvable if and only if the NCL graph was. The maze is planar by its construction, using the planarity of the NCL graph.

  This completes the proof of PSPACE-hardness. Containment in PSPACE is by Lemma~\ref{lem:all gadgets PSPACE}, so the problem is PSPACE-complete.
\end{proof}

\section{2-Player Unbounded Motion Planning}
\label{sec:2-Player Unbounded}

In this section, we analyze 2-player motion planning games with $k$-tunnel reversible deterministic gadgets. We show that any such game which includes an interacting-tunnels gadget is EXPTIME-complete. We do so by a reduction from 2-player unbounded constraint logic, allowing us to reuse some of the work in the prior section. In addition to building the single player AND and OR vertices, we show how to adapt the gadgets to allow different players to have control of different edges. We also build up the needed infrastructure to enforce turn taking in the simulated game.

The construction of crossovers using interacting-$k$-tunnel reversible deterministic gadgets with two states should allow one to show hardness for the planar version of this problem with those gadgets and any others that simulate them. Care must be taken with the layout, timing, and interaction between crossovers so we do not go on to prove such a result in this paper. Unfortunately, the crossover created by the locking 2-toggle in Section~\ref{sec:1-Player Unbounded Planar} does not suffice and thus leaves the question partially open. In addition, the question of noninteracting-$k$-tunnels reversible deterministic gadgets has not been resolved. We are not able to show problems with such gadgets are easy, and Section~\ref{sec:2-Player Bounded} suggests they should be at least PSPACE-hard.

\begin{lemma}\label{lem:2p in EXP}
  2-player motion planning with any set of gadgets is in EXPTIME.
\end{lemma}
\begin{proof}
  A configuration of the maze consists of the state of each gadget and the location of the robot, and has polynomial length. There is a polynomial-space alternating Turing machine which nondeterministically guesses moves for each player and keeps track of the configuration, using existential quantifiers for player 1 and universal quantifiers for player 2. This Turing machine accepts exactly when player 1 has a forced win. Thus the problem is in APSPACE = EXPTIME.
\end{proof}

\begin{figure}
  \centering
  \includegraphics[width=\textwidth]{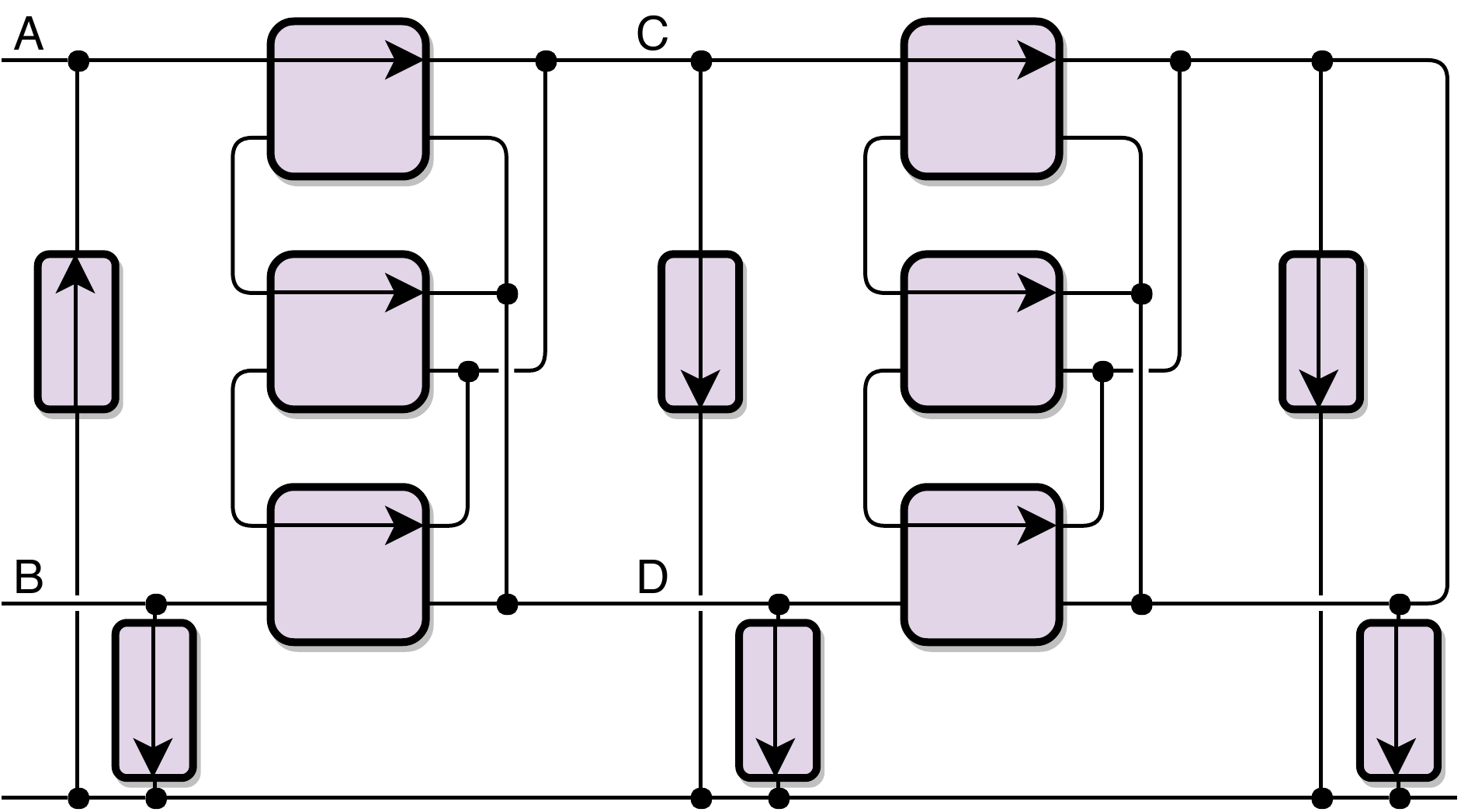}
  \caption{The timer gadget used in the 2CL reduction, made of PL2Ts and 1-toggles. In order to travel between A and B, a player must travel between C and D three times. The timer can be extended to the right; two iterations are shown.}
  \label{fig:2p timer}
\end{figure}

\begin{theorem}\label{thm:2P L2T EXPTIME-c}
  2-player motion planning with the locking 2-toggle gadget is EXPTIME-complete.
\end{theorem}

\begin{figure}
  \centering
  \includegraphics[scale=0.75]{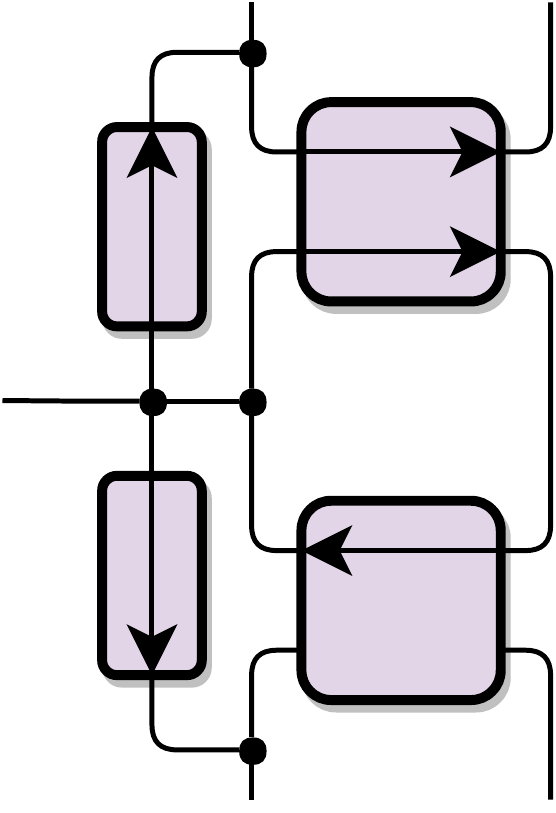}
  \caption{A modified edge gadget for the 2CL reduction. A player can visit the vertex gadgets attached to the edge gadget, and then return to the edge gadget.}
  \label{fig:2p edge}
\end{figure}

\begin{figure}
  \centering
  \includegraphics[width=\textwidth]{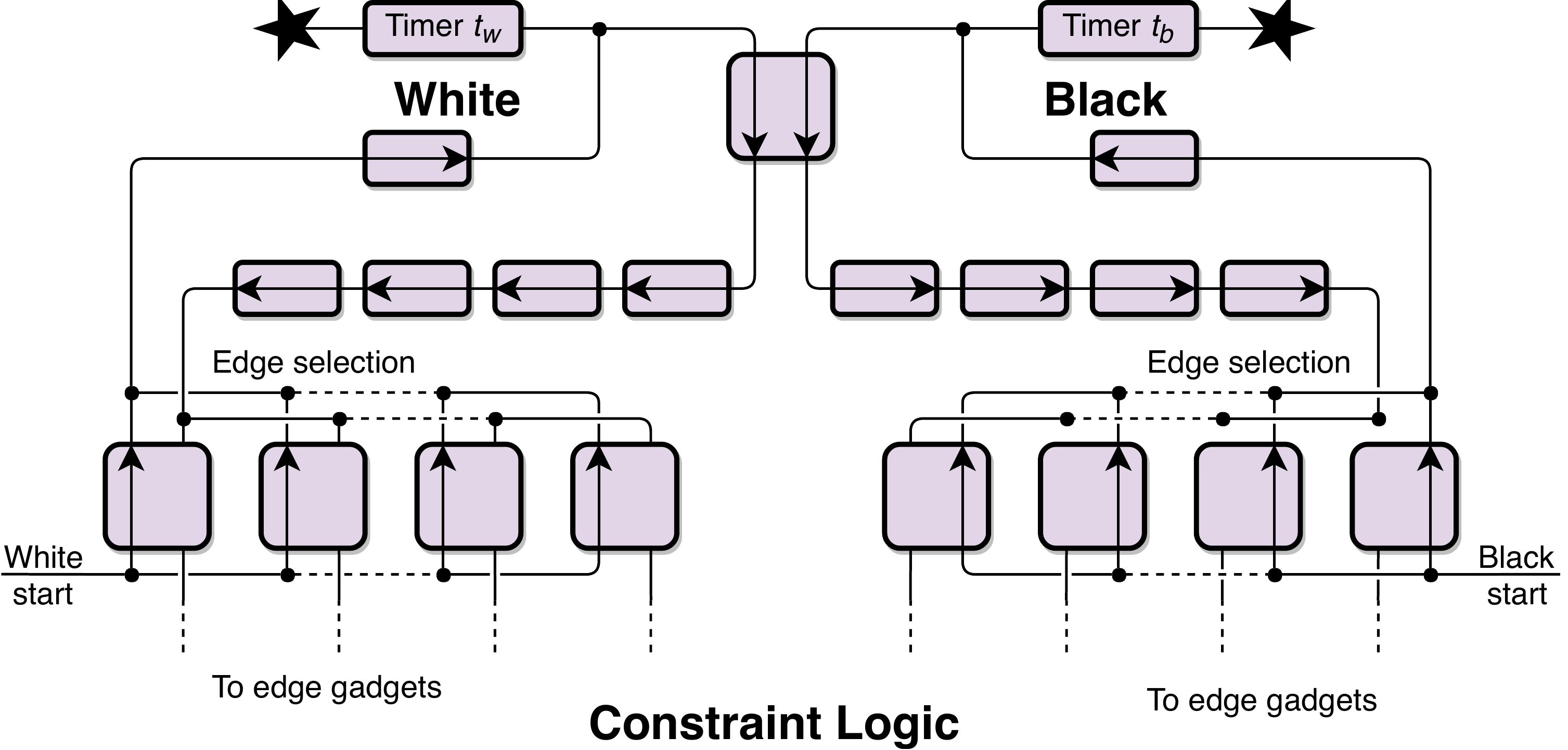}
  \caption{The overall structure and turn enforcement gadget. Each player's edge selection area has a L2T for each edge that player can flip; four are shown for each player. The bottom line from each such L2T connects to the corresponding edge gadget. The timers are as shown in Figure~\ref{fig:2p timer}, with $t_w$ and $t_b$ repetitions. The inside connection to each timer is connected to its access line, and the outside connection (to a win gadget) is at $B$ in Figure~\ref{fig:2p timer}. The goal location past each timer is for the player whose side it is on.}
  \label{fig:2p turns}
\end{figure}

\begin{proof}
  This game is in EXPTIME by Lemma~\ref{lem:2p in EXP}. We use a reduction from 2-player Constraint Logic (2CL) to show EXPTIME-completeness.
  See Appendix~\ref{app:constraint logic} for a definition of 2CL.

  We begin by describing a timer gadget, shown in Figure~\ref{fig:2p timer}. Suppose one player has access to the bottom line. They can enter the gadget at A, and begin going through the timer, eventually reaching a victory gadget at B. The timer has two key properties:

  \begin{enumerate}
    \item Reaching B takes a number of transitions exponential in the size of the timer. In order to get from A to B, the player goes though the top PL2T to C, recursively travels from C to D, goes around the loop through the top two PL2Ts, goes back from D to C, traverses the bottom loop, once again goes from C to D, and finally proceeds to B. If traveling between C and D takes $m$ transitions, then traveling between A and B takes $3m+6$ transitions. If the timer gadget is repeated $k$ times, it takes at least $3^k$ transition to get from A to B.
    \item A player in the timer has an opportunity to exit the timer at least every 2 turns, and exiting takes 1 turn; in particular, they can always exit within 3 turns while progressing the timer. The player uses a 1-toggle to exit to the bottom line. They can then later reenter using the same 1-toggle, resuming their work on the timer where they left off. If the player is in the timer, the next step in progressing the timer is either traversing a loop between to PL2Ts, which takes 2 transitions, or moving horizontally between timer segments, which takes 1 transition. Thus in 3 transition, the player can complete the current or next step and exit to the bottom line.
  \end{enumerate}

  The constraint logic gadgets are similar to those used in Theorem~\ref{thm:L2T PSPACE-c} for the 1-player game, with the modification shown in Figure~\ref{fig:2p edge}. We have added 1-toggles allowing a player at an edge to visit and configure the incident vertices, without allowing the player to travel to other edges. Each player's goal location is inside the gadget corresponding to their target edge, so that they can reach it if they can flip the edge.

  Unlike the 1-player version, we need gadgets to enforce the turn order. The overall construction is shown in Figure~\ref{fig:2p turns}. The maze consists of three main regions: the White area, the Black area, and the constraint logic. Each player will spend most of their time in their own area, occasionally entering the constraint logic to flip an edge. The players' areas are designed to enforce turn order and progression of the game. A player can never enter the other player's area.

  There is a single L2T separating the constraint logic area from each player's area. This prevents both players from being in the constraint logic at the same time.

  Each player's area contains an edge selection gadget, which consist of a locking 2-toggle for each edge they can control. The other line in the L2T is accessible by entering the constraint logic area and passing through a delay line four 1-toggles, and is connected to the corresponding edge gadget. In order to access an edge gadget, the player must activate the appropriate L2T, which requires deactivating the previously activated L2T. This ensures that only one edge gadget is accessible by each player at any time. There is a 1-toggle separating the edge selection gadget from the rest of the player's area, so that switching the selected edge requires at least $4$ turns (we use one tunnel of a L2T for a 1-toggle). 

  Each player's area has a timer, of length $t_w$ for White and $t_b$ for black. If a player finishes their timer, they win.

  Each player begins inside their edge selection gadget, and White goes first. The game begins with White picking an edge and going to the constraint logic area, while Black goes to their timer.

  A round of normal play proceeds as follows:
  \begin{itemize}
    \item White moves from edge selection to the constraint logic area. Black is currently in their timer.
    \item White enters the constraint logic, walks to their selected edge, and flips it. Black continues working on their timer.
    \item White returns through their constraint logic delay line. Once they pass the first 1-toggle, Black finishes their current step in the timer and exits, moving towards edge selection.
    \item White begins working on their timer. Black selects an edge, enters the constraint logic, and flips the edge.
    \item Black returns through their constraint logic delay line. Once they pass the first 1-toggle, White exits their timer and moves to edge selection.
    \item White selects an edge as Black enters their timer.
  \end{itemize}

  Suppose Black has just flipped an edge gadget; they have nothing to do but return through the delay line of length $4$. When Black is past the first 1-toggle, White will leave their timer to flip an edge. Black might try turning around to go back to the constraint logic area. It takes Black at least $6$ turns to flip the edge back, during which White has enough time to select an edge and reenter their timer. The game is now in the same situation as before, except that White has progressed their timer; thus Black does not want to do this.

  Black might instead try waiting at the central L2T after White has selected an edge. White will then go to their timer, forcing Black to exit eventually. When Black is not next to the central L2T, White exits their timer and moves to constraint logic. Because of the 1-toggle separating edge selection from the central L2T, for Black to change their selected edge, they must spend multiple turns away from the L2T, allowing White to enter constraint logic; similarly if Black works on their timer, White can enter constraint logic. So Black has no choice but to pass the turn to White.

  Since White can always exit their timer within 3 turns, and Black has three more 1-toggles to get through when White begins looking to exit, White will reach edge selection before Black can reach edge selection, so White will be the first player ready to enter constraint logic again. Nothing Black can do will prevent White from taking the next turn in the 2CL game. Similarly after White flips an edge, Black will be able to take a turn next. So either player can force the alternation of constraint logic turns.

  The sizes of the timers are chosen to satisfy the following. First, if White cannot win the constraint logic, Black should win, so Black's timer is shorter: $t_b<t_w$. Second, if White can win the constraint logic game, White should win first, even if Black ignores the constraint logic game and just works on their timer. If the constraint logic graph has $n$ edges, it takes at most $2^n$ constraint logic turns for White to win. Each constraint logic turn for White takes $6$ turns to select an edge and return to the constraint logic, $8$ turns to cross the constraint logic delay line twice, $4$ turns to access and flip an edge, and up to $5$ turns to access and configure an incident vertex, so $25$ turns in total during which Black can work on their timer. Both players might be in their timers simultaneously at most $4$ times each cycle, and each time for at most $4$ turns, so Black spends at most $41$ turns in their timer for each constraint logic turn. Thus, since it takes Black at least $3^{t_b}$ turns to win through the timer, we need $41\cdot2^n<3^{t_b}$; $t_b=n+6$ suffices, and we can set $t_w=2n+12$.

  Using these timer sizes, it is clear that the constraint logic game will resolve before either timer if the players follow normal play. We need the timers so that Black cannot force a draw by sitting in the constraint logic forever, preventing White from winning; White will progress on their timer if Black attempts this.

  Hence White has a forced win in the motion planning game if and only if they have a forced win in the constraint logic game. Since 2CL is EXPTIME-complete, the 2-player game on systems of locking 2-toggles is EXPTIME-hard. The maze used in the reduction has only $O(n)$ L2Ts.
\end{proof}

\begin{theorem}
  2-player motion planning with any interacting-$k$-tunnel reversible
  deterministic gadget is EXPTIME-complete.
\end{theorem}

\begin{figure}
  \centering
  \includegraphics[scale=.75]{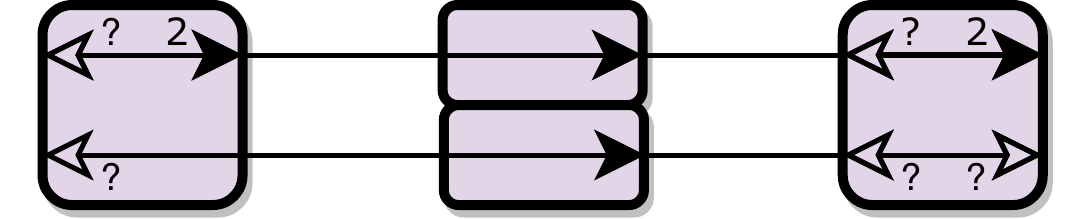}
  \caption{Another state of the construction shown in Figure~\ref{fig:arb gadget sim L2T}. The leftmost gadget is in state 1, and the rightmost gadget is in state 3.}
  \label{fig:arb gadget sim L2T state 1}
\end{figure}

\begin{proof}\label{thm:2P arb EXPTIME-c}
  This game is in EXPTIME by Lemma~\ref{lem:2p in EXP}. We adapt the 2CL reduction in the proof of Theorem~\ref{thm:2P L2T EXPTIME-c}. Replace each locking 2-toggle in that 2CL reduction with the simulation of a locking 2-toggle from the arbitrary gadget in Theorem~\ref{thm:arb gadget sim L2T}. In the new maze, each tunnel in a simulated L2T takes 6 transitions to traverse, so the game goes 6 times slower.

  The simulation still works with two players, as long as both players do not have access to the gadget at the same time. Each L2T in the turn enforcement area is accessible only by one player, and only one player can be in the constraint logic area at any time. The only L2T both players have simultaneous access to is the central gadget which gives access to the constraint logic area, so we look more carefully at that gadget. 

  The state with both edges traversable is shown in Figure~\ref{fig:arb gadget sim L2T} (the 1-toggle simulation still works). Note that the simulation is of an APL2T, but the gadget in the 2CL reduction is a PL2T; this is not a problem because we are not concerned with planarity. Suppose both players approach the gadget, one from the right on the top line and one from the left on the bottom line. Whoever reaches the gadget first should `win the race,' and lock out the other player. The simulation implements this correctly, provided that the player who arrives first is a full turn ahead in the L2T maze, or 6 turns ahead in the new maze. The only time the players might be within 6 turns of each other is at the very beginning of the game, so we put a delay of 6 turns for Black to get from their start location to edge selection to ensure White wins the race by 6 turns. If a player would arrive less than 6 turn before the other player, they should go to their timer instead; since this is a zero-sum game and the players would have to collaborate to break the simulation, one player will choose not to. 

  The other way players can interact at this gadget is when one player is exiting the constraint logic area, and the other player is waiting just outside and enters as soon as they can. The state of the simulation is shown in Figure~\ref{fig:arb gadget sim L2T state 1} (the other possible state is symmetric). One player, say White, has traversed the top edge to enter the constraint logic area, and is about to exit by traversing the top line to the right. Black is waiting at the left end of the bottom line, ready to enter the constraint logic area. The leftmost gadget prevents Black from making any transitions until White begins exiting. Once White begins exiting, the leftmost gadget switches to state 2, so Black can follow parallel to White and one turn behind. As long as White continues through the construction at full speed, Black interacts with the construction as though White has already finished their traversal, so it correctly simulates a L2T. Again breaking the simulation would require the players to cooperate, and the game is zero-sum, so at least one player will ensure the simulation works.
\end{proof}

\section{Team Unbounded}
\label{sec:Team Unbounded}

In this section, we show that team imperfect information games with
interacting-$k$-tunnel reversible deterministic gadgets is
RE-complete, implying the problem is Undecidable.  The reduction is
from Team Private Constraint Logic (TPCL); see
Appendix~\ref{app:constraint logic} for a definition.  We use many of
the ideas and constructions from Section~\ref{sec:2-Player Unbounded},
but various modifications are needed to deal with the additional
player and the model of player knowledge. Recall in this model we have
three players on two different teams, each controlling a single
robot. All players start knowing the configuration of the entire game;
however, after that point players can only observe the states of the
gadgets that their robots can reach via the connection
graph. Adaptations for the planar version and the complexity of such
games with noninteracting-tunnel gadgets remains open as in
Section~\ref{sec:2-Player Unbounded}.

\begin{lemma}\label{lem:team unbounded in re}
Team motion planning with any set of gadgets is in RE
(recursively enumerable).
\end{lemma}

\begin{proof}
Suppose the White team has a forced win on some system of gadgets, and consider the tree of possible positions when White follows their winning strategy. The branches in the tree correspond to choices the Black team might make. Since White forces a win, every branch of the tree is finite. Since Black has finitely many choices at each turn, the tree is finitely branching, so by K\H{o}nig's infinity lemma \cite{Koenig-1927}, the tree is finite. In particular, there is a finite bound on the number of turns it takes for White to win, so the winning strategy can be described in a finite amount of space. So there are countably many potential winning strategies, and we can sort them lexicographically.

Given a potential winning strategy, the problem of determining whether it is actually a winning strategy is decidable: an algorithm can explore every choice Black might make, and see whether White always wins. There are only finitely many choices to check because the strategy only describes a finite number of turns.

We use the following algorithm to determine whether White has a forced win. For each potential winning strategy in lexicographic order, check whether it is a winning strategy. If it is, accept. This algorithm accepts whenever White has a forced win, and runs forever otherwise, so it recognizes the games in which White has a forced win.
\end{proof}

Although \cite{CL_Complexity2008} only mentions undecidability and not RE-completeness, it follows that TPCL is RE-complete. Containment in RE is given by an argument nearly identical to the proof of Lemma~\ref{lem:team unbounded in re}. The proof of undecidability is ultimately by a reduction from acceptance of a Turing machine on an empty input, which is RE-complete, implying that TPCL is RE-hard.

\begin{theorem}\label{thm:team L2T undecidable}
  Team motion planning with the locking 2-toggle gadget is RE-complete
  (and thus undecidable).
\end{theorem}

\begin{figure}
  \centering
  \begin{subfigure}{0.47\linewidth}
    \centering
    \includegraphics[scale=0.5]{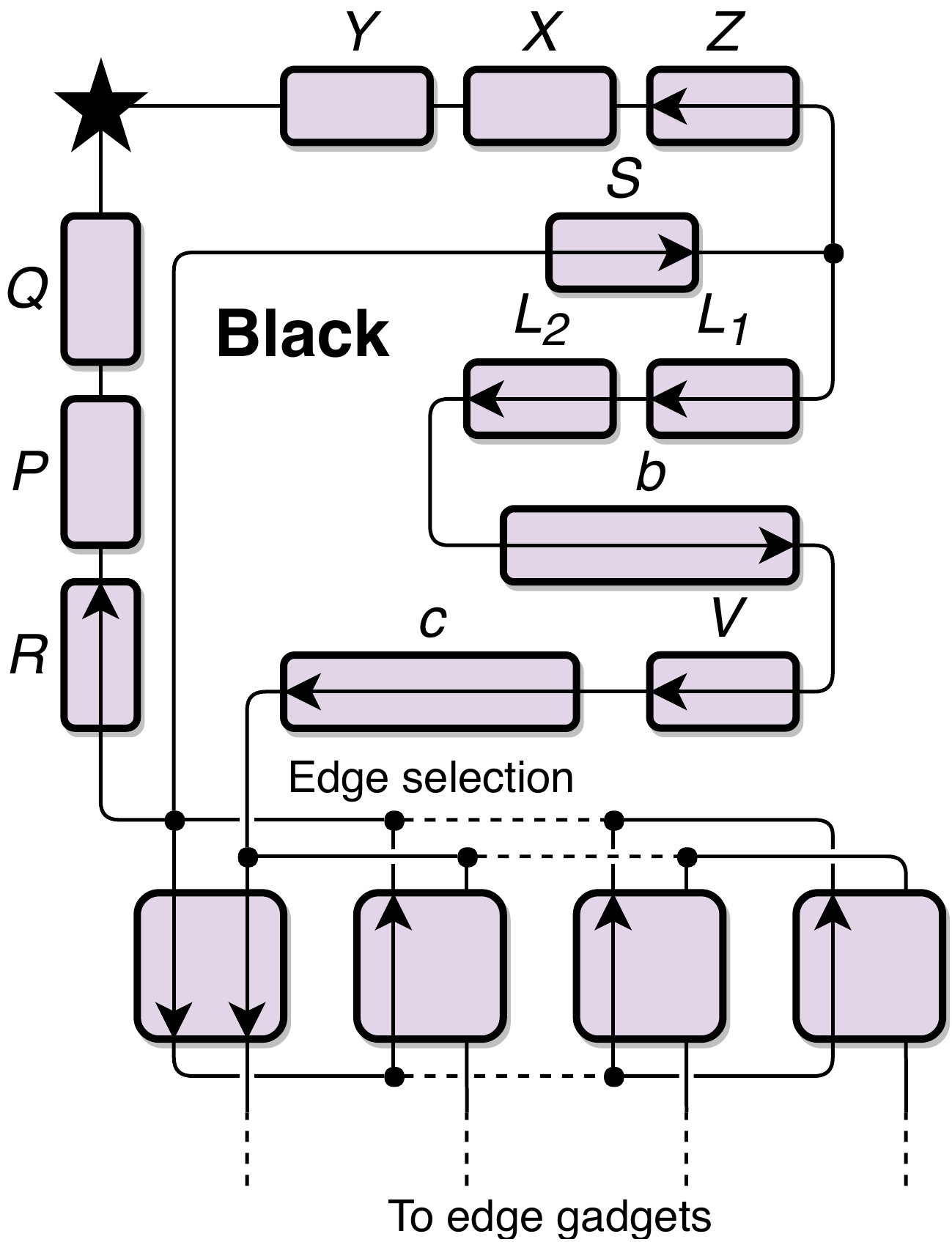}
    \label{fig:team black}
  \end{subfigure}\hfil\hfil
  \begin{subfigure}{0.47\linewidth}
    \centering
    \includegraphics[scale=0.5]{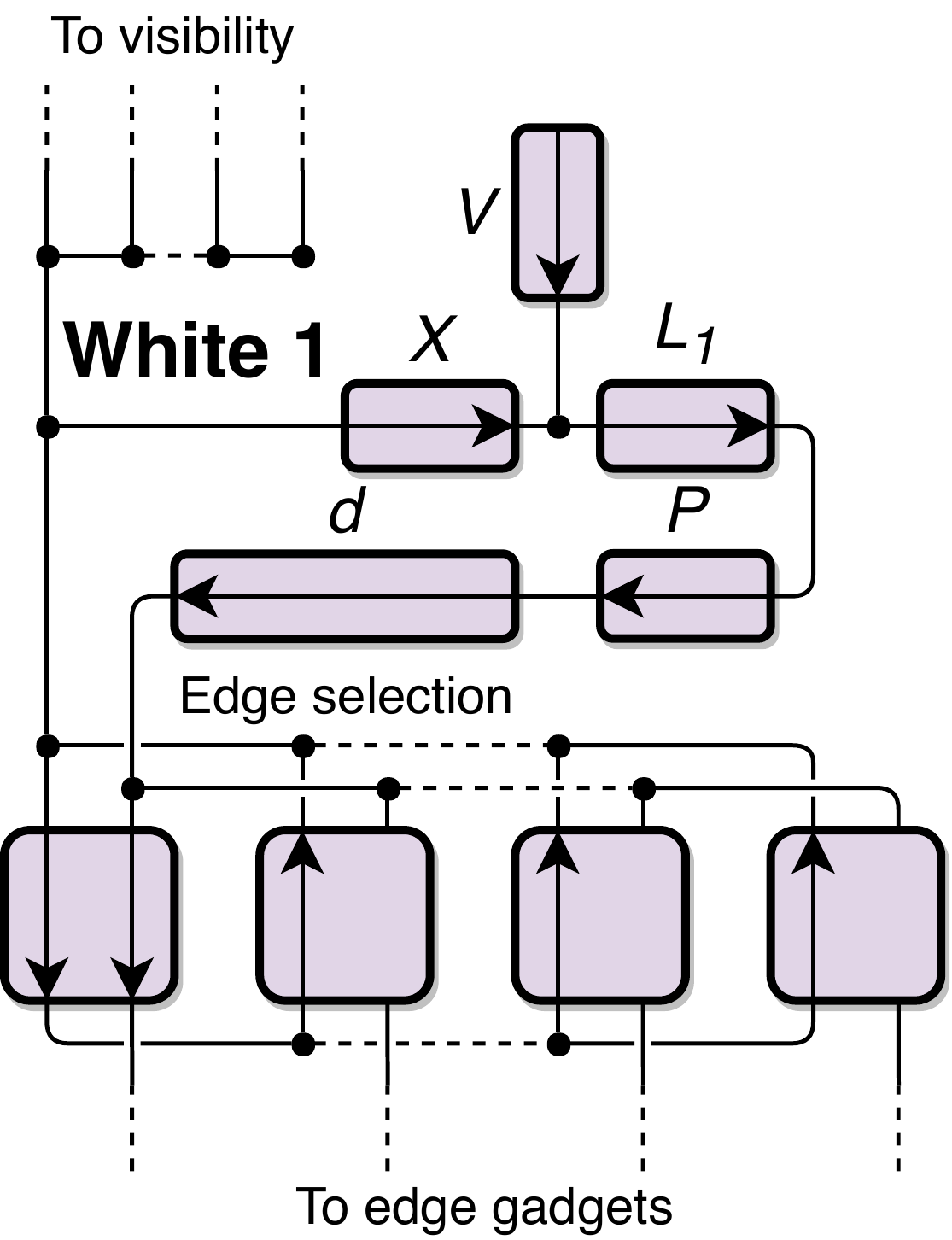}
    \label{fig:team white1}
  \end{subfigure}
  \begin{subfigure}{\linewidth}
    \centering
    \includegraphics[scale=0.5]{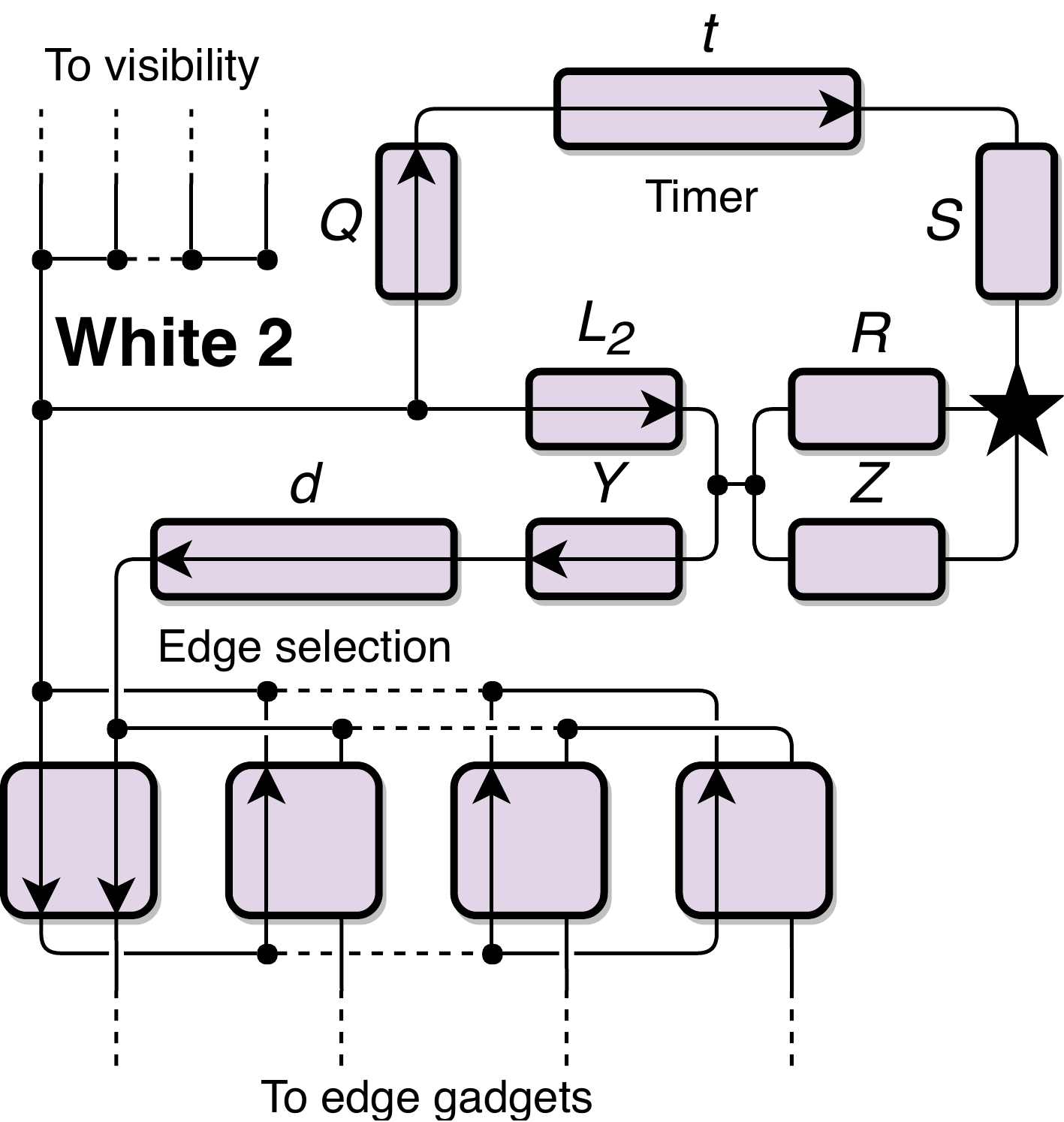}
    \label{fig:team white2}
  \end{subfigure}
  \caption{The turn enforcement gadget for the team game. Each player has their own region which contains an edge selection area, a path to the edge gadgets they can control, and some other constructions. Each White player has a visibility area which allows them to see the state of some edge gadgets in constant time. There is no good layout for the whole gadget, so we use pairs of 1-toggles that share a (capital) label to represent L2T. Long boxes with lowercase labels represent chains of 1-toggles with length given by the label. The win gadgets are for the obvious players, and the tunnels currently not traversable ($P$, $Q$, $R$, $S$, $X$, $Y$, and $Z$) will directed toward the win gadget when they become traversable.}
  \label{fig:team turns}
\end{figure}

\begin{figure}
  \centering
  \includegraphics[scale=.75]{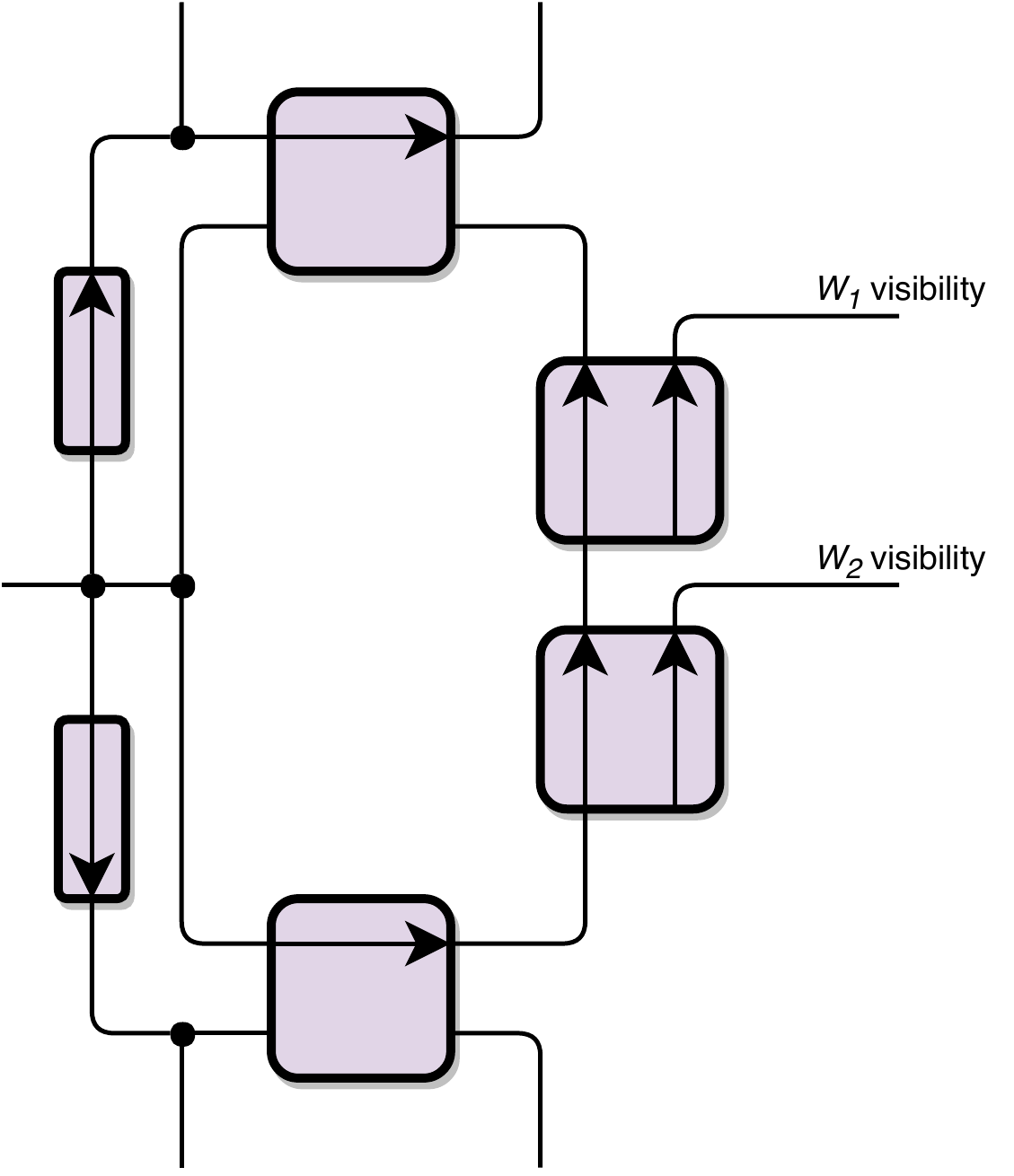}
  \caption{An edge gadget for the TPCL reduction. This is the same as a 2CL edge gadget, except two L2Ts have been added that allow $W_1$ or $W_2$ to see the state of the edge if it is connected to their visibility area, but they cannot make any transitions.}
  \label{fig:team edge}
\end{figure}

\begin{proof}
  Containment in RE is given by Lemma~\ref{lem:team unbounded in re}. For RE-hardness, we use a reduction from TPCL, with a similar construction as in the proof of Theorem~\ref{thm:2P L2T EXPTIME-c}. The overall construction is shown in Figure~\ref{fig:team turns}.  Capital letters label L2Ts, and lowercase letters label lengths of delay lines. The two tunnels in the same L2T are labelled the same, instead of being positioned next to each other. The three players $B$, $W_1$, and $W_2$ each have their own region. Each region contains an edge selection area with $k$ edges initially active, access to the constraint logic, and some additional gadgets. We need to ensure the following:
  \begin{enumerate}
    \item Turn order is enforced. That is, the players take turns in the order $B$, $W_1$, $W_2$, and neither team can gain anything by deviating from this. We use $L_1$ and $L_2$ to prevent $B$ from being in the constraint logic area at the same time as $W_1$ or $W_2$, and appropriate delays to ensure each player is ready for their turn. The timer in $W_2$'s region forces $B$ to eventually pass the turn to $W_1$.
    \item Each player can flip up to $k$ edges each turn. If $k$ edges are initially accessible for each player, the edge selection area allows them to select any $k$ of their edges, and a player must end their turn in order to change their selection.
    \item The White players have the correct information about the state of the game. Each of them has a visibility area, which allows them to see the orientation of the appropriate constraint logic edges. We must not allow $W_1$ and $W_2$ to both access the same L2T, as they could then use it to communicate. So we need a more complicated mechanism to prevent both White players from being to the constraint logic area at the same time.
  \end{enumerate}
  For visibility, we modify the edge gadget as shown in Figure~\ref{fig:team edge}. The appropriate line is connected to each White player's visibility area if they should be able to see that edge.

  A round of normal play proceeds as follows:
  \begin{itemize}
    \item $B$ begins their turn by passing down through $L_1$ and $L_2$. $W_1$ waits next to $V$, and $W_2$ walks through their timer.
    \item $B$ flips some edges, and returns, passing $V$. When $W_1$ sees this happen, they go to their visibility area, and then select $k$ edges. $W_2$ continues in the timer.
    \item $B$ finishes exiting through the delay $b$. Once $B$ has passed $L_1$, $W_1$ enters the constraint logic area. $W_2$ reaches the end of the timer, finds $S$ to be closed, and comes back.
    \item $B$ is stuck on the side of $L_1$ away from the constraint logic area, and can select edges. $W_1$ flips edges and returns to just below $L_1$. $W_2$ goes to their visibility area, and then selects edges.
    \item After a number of turns large enough that both White players are definitely ready, $W_1$ exits $L_1$. The same round, $W_2$ enters $L_2$, passing the turns from $W_1$ to $W_2$.
    \item $W_2$ takes their turn. $B$ waits just to the right of $L_2$, and $W_1$ waits above $X$.
    \item $W_2$ exits $L_2$ and goes to the timer. $B$ passes through $L_2$ to take their turn, and $W_1$ waits.
  \end{itemize}

  We place each player's starting location to be at the end of a chain of 1-toggles leading to their region, so they arrive after an appropriate delay. We can set $B$ to have no delay and $W_1$ and $W_2$ to have $2k$ delay, so $B$ has time to select edges before the White players arrive. The first turn has slightly strange timing since $W_2$ starts the timer later than normal, but this is not important.

  We consider ways in which player might deviate from normal play, and see that in each case they do not gain anything by deviating.

  $B$ enters the constraint logic through $L_2$ as soon as $W_2$ passes $L_2$ on their way out, at which point $W_2$ enters the timer. $B$ need to be able to take a full turn and go back through $S$ before $W_2$ reaches the end of the timer; this takes up to $2(b+c+2)+2+11k$ turns, since flipping each edge now takes up to $11$ turns. So we need $t>2(b+c+2)+2+11k$. The timer forces $B$ to return through $S$ within $t+2$ turns, since otherwise $W_2$ wins. 

  The gadget $V$ lets $W_1$ know when $B$ is done, since $W_1$ can see whether $B$ is past $V$ while waiting at $L_1$. Specifically, $W_1$ waits until they see $B$ stay past $V$ for $2c$ turns, and then return. For $B$ to be unable to flip edges after this, we need $4c>t$. Then $W_1$ goes to visibility and sees the current configuration, selects $k$ edges for their next turn, and waits at $L_1$ again. For $W_1$ to have time to do this before $B$ gets out, we need $b>2k+2$. 

  Once $B$ exits $L_1$, $W_1$ goes in and flips edges. The delay $d$ ensures that if $W_1$ (or $W_2$) flips any edges, then $B$ will be ready for their next turn; we need $2d>2k+4$. $W_2$ returns through the timer, checks visibility, and selects edges. If $W_2$ enters constraint logic before $W_1$ leaves, $B$ can win through $X$ and $Y$, so $W_2$ must wait until $W_1$ leaves. The White players coordinate using the fact that the length of an entire round is bounded, so they can wait long enough to ensure that they are both ready, and then $W_1$ exits $X$ immediately before $W_2$ enters $Y$. Since $W_1$ was past $L_1$, $B$ is locked outside of $L_1$, so $W_2$ can get past $L_2$; the $W_1$ can safely pass the turn to $W_2$.

  While $W_1$ is past $X$, $B$ might try going through $Z$ and $X$, trapping $W_1$. In this case, $W_2$ can win through $Z$, so $B$ will only go through $Z$ if both $X$ and $Y$ are traversable.

  During $W_2$'s turn in the constraint logic, $W_1$ must not be past $X$ to prevent $B$ from winning through $X$ and $Y$. So $B$ can go through $L_1$, and go through $L_2$ as soon as $W_2$ exits. That is, $W_2$ cannot pass the turn back to $W_1$. 

  $W_2$ might try to stay in the timer, forcing $B$ to stay out of the constraint logic to prevent $W_2$ from winning through $S$. Then $W_1$ might be able to take extra turns in the constraint logic. If the White team attempts this, $B$ will win through $P$ and $Q$. If $B$ goes through $R$ and $P$ when $Q$ is not traversable in order to trap $W_1$, $W_2$ will win through $R$; these three L2Ts are analogous to $X$, $Y$, and $Z$.

  Assuming the constraints mentioned are satisfied, no player or team can usefully deviate from normal play, and normal play simulates the TPCL game. Thus White has a forced win in the team motion planning game if and only if they have a forced win in the TPCL game.

  We can satisfy all the constraints, e.g by $b=2k+3$, $c=8k+7$, $d=k+3$, and $t=31k+27$ (the constraints are not tight, but they suffice). The number of L2Ts in the resulting system of gadgets is only linear in the number of edges in the constraint logic graph. 
\end{proof}

\begin{theorem}\label{thm:team arb undecidable}
  Team motion planning with any interacting-$k$-tunnel reversible deterministic
  gadget is RE-complete.
\end{theorem}

\begin{figure}
  \centering
  \includegraphics[scale=.75]{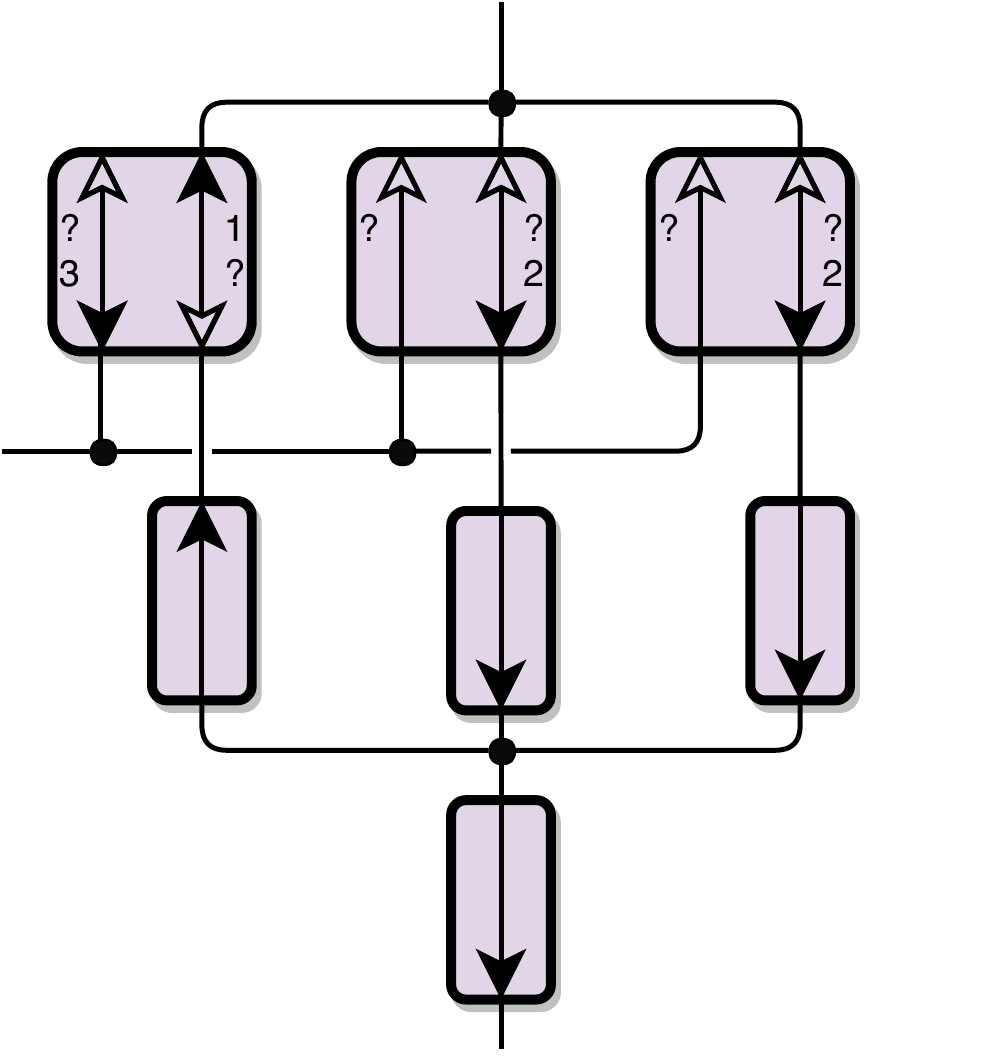}
  \caption{A visibility gadget for the TPCL reduction. The Black player can travel between the top and bottom, and a White player can enter the side to see which direction was traversed most recently.}
  \label{fig:visibility}
\end{figure}

\begin{proof}
  Containment in RE is given by Lemma~\ref{lem:team unbounded in re}. For RE-hardness, we adapt the TPCL reduction in Theorem~\ref{thm:team L2T undecidable} to work for the arbitrary gadget. As in the 2-player case of Theorem~\ref{thm:2P arb EXPTIME-c}, it is almost sufficient to replace each L2T with the simulation in Theorem~\ref{thm:arb gadget sim L2T}. We examine the L2Ts that are shared between two players.

  First, $L_1$ and $L_2$ are analogous to the central L2T in Theorem~\ref{thm:2P arb EXPTIME-c}: if two player are racing to enter, the player who should win is at least 6 turns ahead, and if one player exits and another enters, is works correctly.

  For $S$, $P$, $Q$, $R$, $X$, $Y$, and $Z$, we use a single copy of the arbitrary gadget with 5 extra gadgets for delay, instead of the simulation. Considering the gadget as in Figure~\ref{fig:arb gadget}, we use state 1, and put the bottom edge in the position next to a win gadget. For $S$, $Q$, $Y$, $R$, and $Z$, if the bottom edge is traversed from state 2, the game is over, so the gadget is never in a state other than 1 or 2 while the game is going. For $P$ and $X$, we know that $B$ cannot safely wait past those gadgets, so the game must be about to end in Black victory if they ever reach state 3.

  For $V$ and the visibility gadgets on edges, we use the construction in Figure~\ref{fig:visibility}. $B$ has three paths to choose from in the process of crossing the bottommost 1-toggle, and always two of them are align with that 1-toggle, so $B$ has two options. The White player, say $W_1$ can see the state of a gadget in all three paths, and thus determine the orientation. If $W_1$ goes through one of these gadgets, $B$ will use the other path. If there were only one path, $W_1$ could go through the gadget, forcing $B$ to either not flip that edge or get a gadget into an unknown state (for L2Ts, we used the fact that $W_1$ could never traverse that tunnel in one direction). This visibility gadget allows $W_1$ to see the orientation of a constraint logic edge or $V$ without being able to interfere.  

  Once we make these replacements, the new maze with the arbitrary gadget has a forced win by White if and only if the maze with L2Ts did.
\end{proof}

\section{1-Player Bounded Motion Planning}
\label{sec:1-Player Bounded}
In this section, we consider a broad class of gadgets which are naturally in NP and give a dichotomy classifying them as NP-complete or in NL. We examine all gadgets in tunnels whose state-transition graph forms a DAG. We will call these DAG gadgets for short. Our proof of hardness further applies to a larger class of gadgets, however a full classification of more general, simple to describe classes of gadgets will require more insight or much more case-work. Also, our constructions require the use of a crossover gadget.

The results in this section can be seen as similar to Viglietta's Metatheorem 1 about location traversal (being implemented by the interacting tunnels in gadgets) and single-use paths \cite{HardGames12}. It also bears resemblance to  Metatheorem 4 about pressure plates which only affect one door \cite{HardGames12}. However, our proof goes through 3SAT rather than Hamiltonian Path, uses a different underlying model which makes different features salient, and gives generalizations in a different direction. Structurally the proof follows that used to show Mario as well as many other games are NP-hard \cite{NintendoFUN2014}.

\begin{lemma}\label{lem:DAG has single-use}
All DAG gadgets contain a single-use transition unless they are a transitionless gadget. 
\end{lemma}
\begin{proof}
First, find a node which only has transitions to \emph{terminal states}, ones with no possible further transitions. To find one, begin by removing all terminal states from the graph. Of the remaining nodes, all of them which are now terminal states must have pointed to at least one terminal state in the original graph or it would have been removed, and it must have only pointed to terminal states or it would not be a terminal node. Terminal nodes have no transitions. Thus the node we discovered has an available transition which closes all tunnels in the new state. The gadget starting from that state is a single-use gadget.
\end{proof}

In a system of gadgets, each DAG gadget can only be traversed polynomially many times. This is the core reason that motion planning involving these gadgets is always in NP.

\begin{lemma}\label{lem:DAG in NP}
1-player motion planning with any set of DAG gadgets is in NP.
\end{lemma}
\begin{proof}
If a gadget and its state is a sink in the state-transition graph, then no transitions are available from that state. Each time a gadget is traversed the state of the gadget is moved down the graph. All paths from any vertex to a leaf are of polynomial length and thus each gadget can only be traversed polynomially many times before it no longer has any open tunnels. Thus we can give a polynomial size witness consisting of the order in which gadgets are visited, as well as the transition made at each gadget. To verify this certificate we check that each specified transition is legal, and that the location after each transition is connected to the location before the next transition in the witness.
\end{proof}

Recall from Theorem~\ref{thm:trivial} that all gadgets without interacting tunnels are in NL. Thus one might hope to show that all interacting-$k$-tunnel DAG gadgets are NP-complete. This is true for deterministic gadgets but false in general; nondeterministic gadgets require a more careful categorization. We will define two behaviors a DAG gadget might have, `distant opening' and `forced distant closing,' and show that either behavior guarantees NP-hardness, while having neither one puts the gadget in NL.

A \emph{distant opening} in a DAG gadget is a transition in some state across a tunnel which opens a different tunnel.

\begin{lemma}\label{lem:open np}
1-player motion planning with any $k$-tunnel DAG gadget with a distant opening is NP-hard.
\end{lemma}
\begin{proof}
We show this problem is hard by a standard reduction from 3SAT. See Appendix~\ref{app:formula games} for a definition of 3SAT.

We construct our reduction as follows. We use the tunnel which is traversed in the distant opening and one of the tunnels it opens. Each literal in a 3-CNF formula will be represented by those two tunnels in a single gadget, in the state of the distance opening. Each variable $x_i$ is represented by a connection to two different paths, one which goes through the opening transitions for the $x_i$ literals, and one for the $\neg x_i$ literals. We place a single-use gadget at the start and end of each branch of each variable to ensure only one side of the variable is traversed. The single-use gadget prevents the agent from returning on the same branch, and if the agent returns via the other branch, they will not be able to proceed to the next variable. 

Each clause contains connections between the openable tunnels for each of its literals. All variable gadgets are laid out in series followed by the clause gadgets, with the goal location at the end of the clause gadgets. Each clause gadget can only be traversed if at least one of its corresponding variable gadgets has been traversed, allowing at least one passage to be open. The agent can reach the goal location exactly when it has a path through the variable gadgets which makes each clause gadget traversable, which corresponds to a satisfying assignment of the 3-CNF formula.
\end{proof}

When a transition across a tunnel closes another tunnel, the situation is more complicated, since the agent may be able to cross the same tunnel through a different transition, choosing not to close the other tunnel. For distant openings, the agent always chooses to open the other tunnel. We will now consider only \emph{monotonically closing} DAG gadgets, which are DAG gadgets with no distant openings.
We clarify some terminology regarding $k$-tunnel DAG gadgets. A \emph{transition} is an edge in the transition graph, which is a legal move between locations which changes the state of the gadget. A \emph{traversal} in a state is an orientation of a tunnel which is open in that state. A traversal may correspond to multiple transitions; a gadget being deterministic is equivalent to each traversal having only one transition. An \emph{orientation} of a set of tunnels in a state contains, for each tunnel in the set, a single traversal of the tunnel the state.

For NP-completeness one might suggest there exist a traversal such that all of its transitions close some other traversal. However, this fails in a simple two tunnel case where one transition closes one direction of the other tunnel and the other transition closes the other direction. This leads us to a more complex definition. A \emph{forced distant closing} in a state of a DAG gadget is a traversal across a tunnel in that state and an orientation of some other tunnels in the state such that, for each transition corresponding to the traversal, the transition closes some traversal in the orientation. The \emph{size} of a forced distant closing is the number of traversals in the orientation.

\begin{lemma}\label{lem:close np}
1-player motion planning with any monotonic $k$-tunnel DAG gadget with a forced distant closing is NP-hard.
\end{lemma}

\begin{proof}
Consider all states which have forced distant closings, and let $s$ be such a state that is minimal in the state-transition DAG, so that after making a transition from state $s$ there are no forced distant closings. We will use a forced distant closing in $s$ with smallest size; say this forced distant closing traverses tunnel $t$ and has size $i$. We chain the $i$ tunnels in the orientation for the forced distant closing, in the directions specified by the orientation, to make what is effectively a single long tunnel $r$. We will use the tunnels $t$ and $r$ in a reduction from 3SAT, and they have two important properties:
\begin{itemize}
  \item If the agent traverses $t$, it cannot later traverse $r$: since we are using a forced distant closing, after traversing $t$ at least one (oriented) tunnel in $r$ is not traversable. Since there are no distant openings, this tunnel cannot become traversable again.
  \item The agent can traverse $r$ from state $s$: in state $s$, each tunnel in $r$ is open. The agent begins by traversing the first tunnel in $r$. This cannot be a forced distant closing for the remaining $i-1$ tunnels, since we assume the smallest forced distant closing has size $i$. So the agent can choose a transition which leaves the remaining tunnels in $r$ open. After this first traversal, there are no more forced distant closings, so the robot can always choose a transition which leaves the remaining tunnels in $r$ open.
\end{itemize}
We can now describe the reduction, which is very similar to the reduction in the proof of Lemma~\ref{lem:open np}. Each literal in a 3-CNF formula is represented by a gadget in state $s$, with the tunnels $r$ chained together. Each variable $x_i$ is represented by a connection to two different paths, one which goes through $t$ for the $x_i$ literals, and one for the $\neg x_i$ literals. We place a single-use gadget at the start and end of each branch of each variable to ensure only one side of the variable is traversed. The single-use gadget prevents the agent from returning on the same branch, and if the agent returns via the other branch, they will not be able to proceed to the next variable. 

When the agent goes through the $x_i$ (resp $\neg x_i$) path of a variable, it closes $r$ in the gadget for each literal $x_i$ ($\neg x_i$), which corresponds to assigning $x_i$ to false (true). This is reversed from the reduction for gadgets with distant openings.

Each clause contains connections between the $r$ for each of its literals. All variable gadgets are laid out in series followed by the clause gadgets, with the goal location at the end of the clause gadgets. Each clause gadget can only be traversed if at least one of its corresponding variable gadgets has \textit{not} been traversed, leaving at least one passage $r$ open. The agent can reach the goal location exactly when it has a path through the variable gadgets which leaves each clause gadget traversable, which corresponds to a satisfying assignment of the 3-CNF formula.
\end{proof}

\begin{lemma}\label{lem:easy dag}
1-player motion planning with any monotonic $k$-tunnel DAG gadget with no forced distant closing is in NL.
\end{lemma}

\begin{proof}
The proof follows that of Theorem~\ref{thm:trivial}, though we must be more careful to account for optional distant closings. As in Theorem~\ref{thm:trivial}, if a system of gadgets has a solution, then a solution of minimal length does not intersect itself. This only requires that the gadget has no distant openings, since then making transitions can never increase traversability, and the shortcutting argument applies.

We locally convert the system of gadgets into a directed graph, and show a path in the graph from the start location to the goal location corresponds to a solution to the system of gadgets which does not intersect itself. Given a (not self-intersecting) path in the graph, we follow the corresponding path through the system of gadgets. When we make a traversal, we must pick a transition to avoid closing tunnels we will need later. This is always possible because there are no forced distant closings; we can always choose a transition which does not close any traversal in the orientation consisting of the traversals the path will later take. By doing this, we ensure that every traversal we need is available when we get to it, so the system of gadgets is solvable.

Suppose there is a solution to the system of gadgets that does not intersect itself. Since it uses each tunnel at most once, and the gadget has no distant openings, the traversability of each tunnel does not change before the solution uses it. Thus the solution is also a path in the directed graph.

So the system of gadgets has a solution iff there is a path from the start location to the end location in the directed graph. Since we can locally convert the system of gadgets to the graph in logarithmic space and solve reachability in NL, the motion planning problem is in NL.
\end{proof}

Combining Lemmas~\ref{lem:DAG has single-use}, \ref{lem:DAG in NP}, \ref{lem:open np}, \ref{lem:close np}, and \ref{lem:easy dag}, we have our dichotomy:

\begin{theorem}\label{thm:1p bounded dichotomy}
1-player motion planning with a $k$-tunnel DAG gadget is NP-complete if the gadget has a distant opening or forced distant closing, and otherwise is in NL.
\end{theorem}

It is natural to wonder whether this condition for hardness can be checked in polynomial time. That is, is there a polynomial-time algorithm which determines whether 1-player motion planning with a given DAG gadget is NP-complete? For all of our other dichotomies, the question of whether a gadget of the appropriate type satisfies the condition for hardness is clearly in P; in fact, in L. But a forced distant closing involves an orientation of the tunnels in the gadget, so there may be exponentially many potential forced distant closings to check. We will show that whenever it is necessary to search through each potential forced distant closing, the number of states of the gadget is exponential in the number of tunnels, so the search takes time polynomial in the number of states.

First, it is easy to determine whether a DAG gadget has a distant opening in polynomial time, since we can iterate through the transitions and see whether each one opens another tunnel. So we consider gadgets with no distant openings, and wish to determine whether they have a forced distant closing.

\begin{lemma}\label{lem:many states}
  Suppose a monotonic DAG gadget has a state $s$ with $k$ open tunnels, and there are no forced distant closings from states reachable from $s$. Then the gadget has at least $2^k$ states reachable from $s$.
\end{lemma}

\begin{proof}
  For each subset of the open tunnels in $s$, we will find a state that has exactly those tunnels open. Since there are $2^k$ such subsets, this implies there are at least $2^k$ states. Assume without loss of generality that each tunnel is traversable from left to right in state $s$.

  Given a subset $X$ of the open tunnels, we perform transitions starting from $s$ as follows. For each tunnel not in $X$, traverse the tunnel repeatedly until it is closed in both directions; this must happen eventually because the gadget is a DAG. At each traversal, choose a transition which does not close any other tunnel from left to right. If there were no such choice of transition, that traversal with all other tunnels oriented from left to right would be a forced distant closing, which does not exist by assumption.

  After making these transitions, we have closed each tunnel not in $X$ without closing any tunnels in $X$. Since the gadget is monotonic, we have not reopened any tunnel. So the final state has exactly the tunnels in $X$ open.
\end{proof}

\begin{theorem}
Deciding whether a 1-player motion planning with a $k$-tunnel DAG gadget is NP-complete can be done in polynomial time.
\end{theorem}
\begin{proof}

The following algorithm checks in polynomial time whether 1-player motion planning with a given a DAG gadget is NP-complete.

\begin{itemize}
  \item For each transition, see whether it is a distant opening. If it is, accept.
  \item Iterate through the states of the gadget in reverse order; i.e. check each state reachable from $s$ before checking $s$. For each state, and for each traversal from that state:
  \begin{itemize}
    \item Suppose the state has $k$ open tunnels other than the tunnel of the traversal. If every transition corresponding to the traversal leaves fewer than $k$ of these tunnels open, accept.
    \item Enumerate the $2^k$ orientations of these $k$ open tunnels, and check for each orientation whether it is a forced distant closing with the traversal. If it is, accept.
  \end{itemize}
  \item Reject.
\end{itemize}

If the gadget has a distant opening, the algorithm notices it in the first step. Otherwise, we check for each state and traversal whether it has a forced distant closing. If every transition for a traversal reduces the number of other open tunnels, than any orientation of the other tunnels gives a forced distant closing. Otherwise, we check for each orientation whether it gives a forced distant closing. So the algorithm accepts exactly when the gadget has a distant opening or a forced distant closing, which is when 1-player motion planning with the gadget is NP-complete by Theorem~\ref{thm:1p bounded dichotomy}.

The only step of the algorithm which does not obviously take polynomial time is running through all $2^k$ orientations of tunnels. Suppose the algorithm reaches this step for some state and traversal. Then there are no forced distant closings after making a transition from this state, since we would have accept already if there were. Also, there is some transition corresponding to the traversal which leaves all $k$ other open tunnels open. By Lemma~\ref{lem:many states}, there are at least $2^k$ states reachable after making this transition. In particular, the gadget has more than $2^k$ states, so enumerating the $2^k$ orientations takes time polynomial in the number of states. Thus the algorithm runs in polynomial time.
\end{proof}

\section{2-Player Bounded Motion Planning}
\label{sec:2-Player Bounded}
In this section, we show that it is PSPACE-complete to decide who wins in a 2-player race with any nontrivial DAG gadget (having at least one transition). To do so we give a construction that shows hardness for single-use paths and single-use one-way gadgets by a reduction from QBF. A simpler construction is possible, but this construction is more easily adapted to the team game in Section~\ref{sec:Team Bounded}. This gives us a nice example of the 2-player local motion planning problem fitting into the canonical complexity class for two-player bounded games. It is also of interest because of how incredibly simple this gadget is. Two-location gadgets trivially do not have interacting tunnels (there is no other tunnel to interact with) and thus the 1-player version of these problems are contained in NL by Theorem~\ref{thm:trivial}.

\begin{lemma}\label{lem:2-player bounded in pspace}
2-player motion planning with any set of DAG gadgets is in PSPACE.
\end{lemma}
\begin{proof}
Since each gadget can undergo only a polynomial number of transitions, the length of the game is polynomially bounded. An alternating Turing machine which uses $\forall$ states to pick Black's moves and $\exists$ state to pick White's moves can simulate the game in polynomial time, so the motion planning problem is in AP${}={}$PSPACE.
\end{proof}

\begin{figure}
  \centering
  \includegraphics[width=\linewidth]{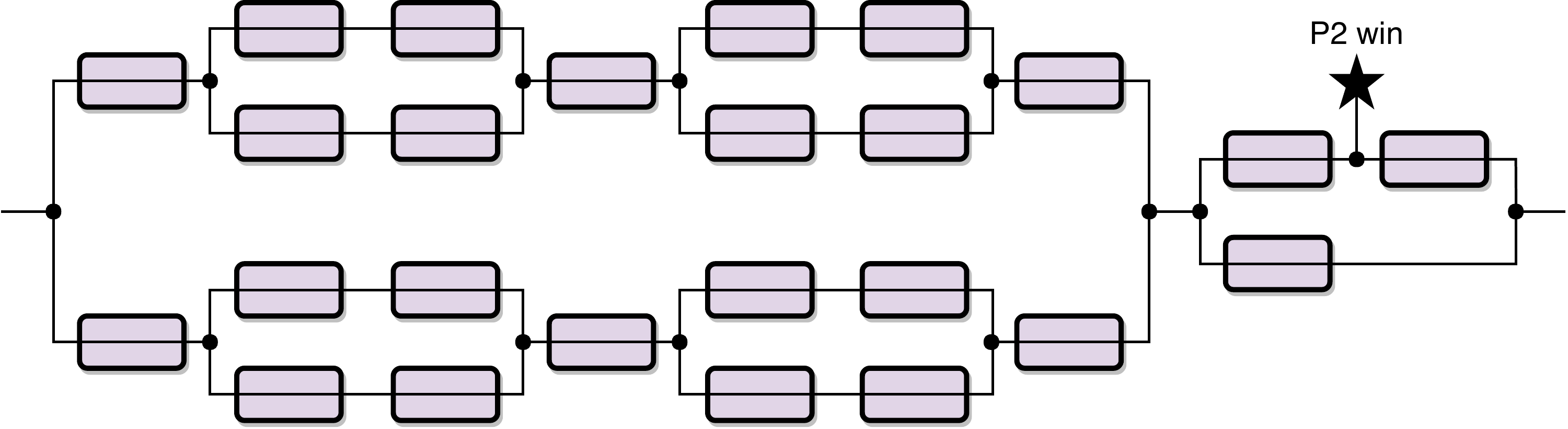}
  \caption{A variable gadget. The players arrive at the left, each take one path across, and exit at the right.}
  \label{fig:2p bounded var}
\end{figure}

\begin{figure}
  \centering
  \includegraphics[scale=0.6]{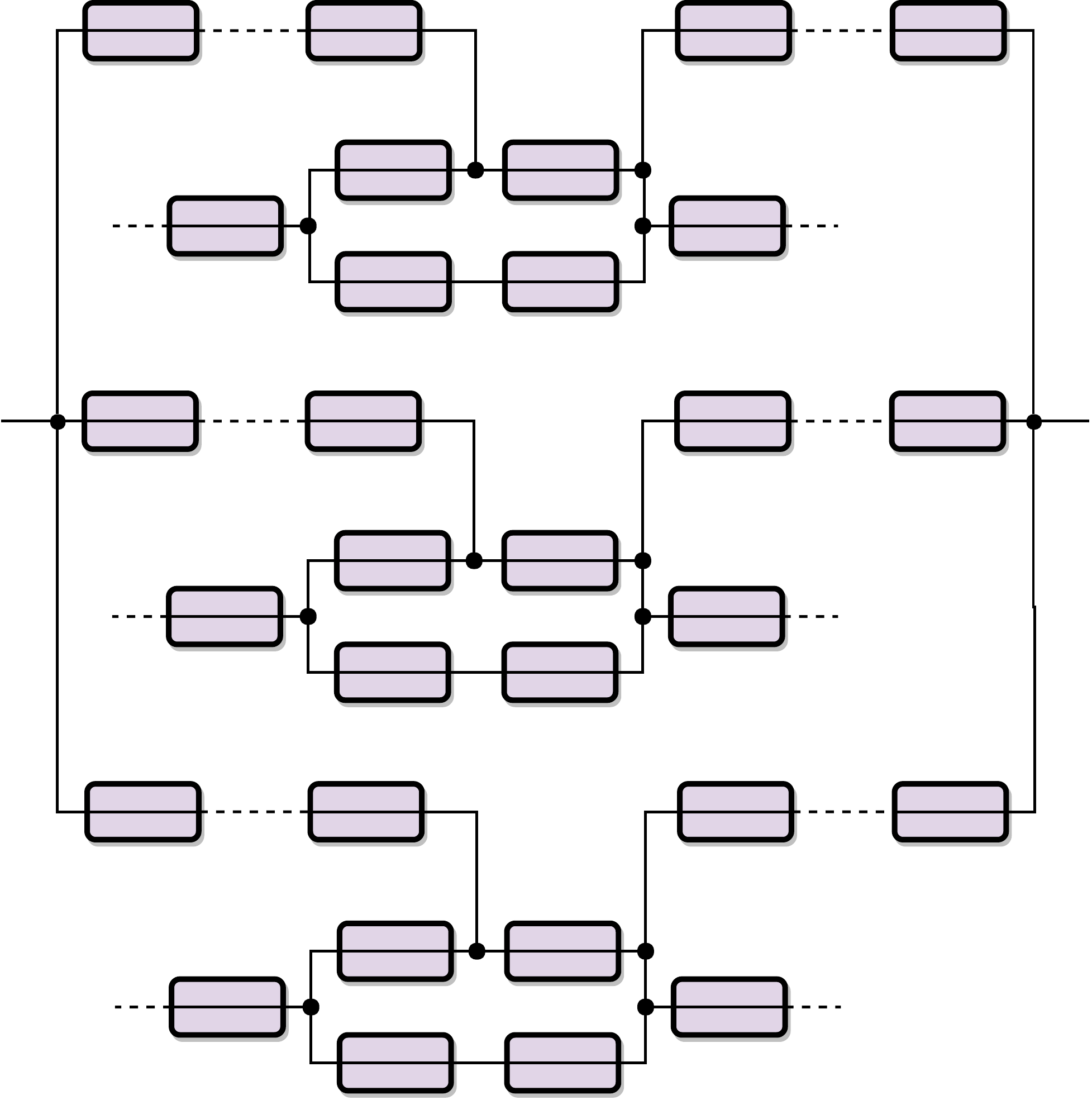}
  \caption{A clause gadget. Each literal is also part of a variable gadget. Each branch has a long series of gadgets so that it takes a large amount of time to traverse.}
  \label{fig:2p bounded clause}
\end{figure}

\begin{figure}
  \centering
  \includegraphics[scale=0.6]{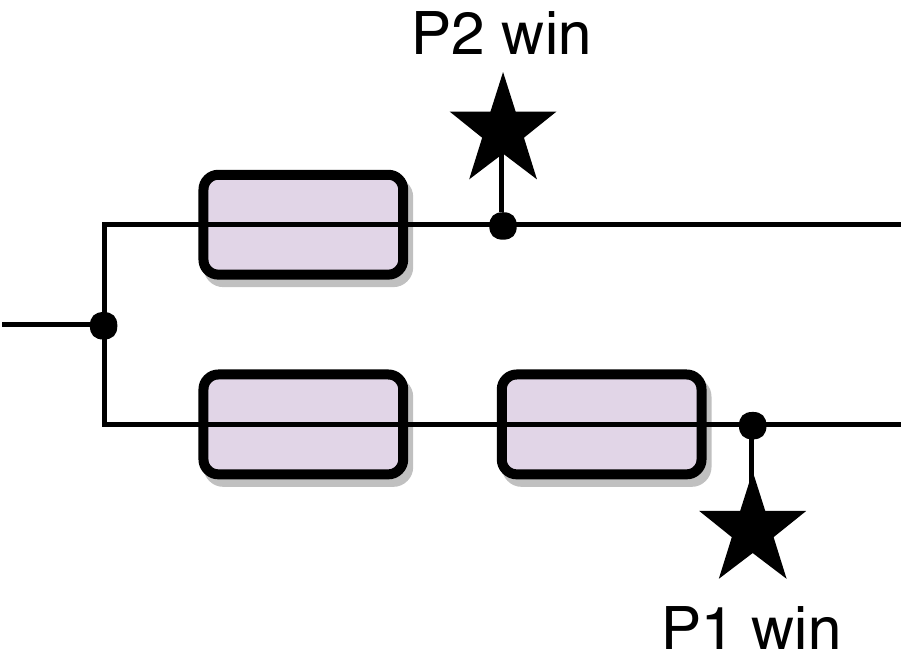}
  \caption{A race gadget. If Player 1 arrives at the left immediately before Player 2, each player ends up on one of the right exits. Otherwise, the player who arrives first wins.}
  \label{fig:2p bounded race}
\end{figure}

\begin{lemma}\label{lem:2-player bounded two-way}
2-player motion planning with the single-use bidirectional gadget is PSPACE-complete.
\end{lemma}
\begin{proof}
Containment in PSPACE follows from Lemma~\ref{lem:2-player bounded in pspace}. For PSPACE-hardness, we reduce from quantified boolean formulas (QBF). See Appendix~\ref{app:formula games} for a definition of QBF. 

We begin by describing the gadgets used in the reduction. The variable gadget is shown in Figure~\ref{fig:2p bounded var}. Most of the gadget is two branches, corresponding to a variable and its negation. Each branch has a series of forks separated by single-use paths. There will be a number of forks depending on the number of occurrences of a literal in the formula; two forks are shown. Each side of each fork has two single-use paths in series. The game will be constructed so that White always prefers the top side of a fork to be traversable, and Black prefers them to be not traversable; the top of a fork will be used later in evaluating the formula.

During the game, both players will pass through each variable gadget, with one player taking each of the two branches. White will take the bottom side of each fork on their branch, and Black will take the top side. Afterwards, only the branch which White took will have forks whose top sides are traversable. Thus we consider the assignment of the variable to be the literal corresponding to the branch White takes.

Suppose both players are at the left end of a variable gadget, and it is Player 1's (who may be White or Black) turn. Player 1 picks a branch, and Player 2 must walk down the other branch. Player 1 arrives at the right end of the branches immediately before Player 2. If Player 1 proceeds along the bottom path, Player 2 wins, so Player 1 must take the top path, which takes one turn longer. After traversing the variable gadget, both players are at the right end, and it is Player 2's turn, so the other player gets to choose a branch in the next variable gadget. 

The clause gadget is shown in Figure~\ref{fig:2p bounded clause}. There are three paths from the left end to the right end, corresponding to the literals in a clause. Each path goes through a fork in a variable gadget. After variables are assigned, the single-use paths on each end of the fork are used, as are either those on the top or those on the bottom of each fork. If the top single-use paths are used, that path through the clause gadget is blocked, and if the bottom paths are used, that path is open. White will ultimately win by traversing each clause gadget, so White prefers to use the bottom side of a fork, and Black prefers to use the top side.

Each path has a large amount of delay (gadgets in series) before and after the fork, so that trying to use the clause gadget during variable assignment results in losing before reaching the end of the delay.

The race gadget is shown in Figure~\ref{fig:2p bounded race}. It ensures both players proceed though variable gadgets as fast as possible. Let Player 1 be the player who reaches the race gadget first in this situation, immediately before Player 2; they are also the player who did not pick the assignment of the last variable. If Player 1 takes the bottom path, Player 2 will win, so Player 1 takes the top path. Then Player 2 takes the bottom path, and now the two players have been separated.

If Player 1 arrives more than a turn ahead of Player 2, they can take the bottom path. The next turn, before Player two can do anything at the race gadget, Player 1 wins. If Player 2 reaches the race gadget first, they can take the top path and win. 

Given a quantified boolean formula with $V$ variables and $C$ clauses, we construct a system of gadgets as follows. We assume the QBF has alternating quantifiers beginning with $\exists$. There is a series of variable gadgets connected end-to-end corresponding to the variables of the formula, in the order of quantification. The goal location inside each variable gadget is a win for alternating players, beginning with Black. The branches of the variable gadget corresponding to $x$ correspond to the literals $x$ and $\neg x$. Each branch of that variable gadget has enough forks that each instance of a $x$ or $\neg x$ in the formula corresponds to a fork, and the two branches have the same number of forks.

There is a clause gadget for each clause in the formula, connected in series. The three branches of a clause gadget correspond to the three literals in the clause. Each branch goes through the fork in the appropriate variable gadget corresponding to that instance of the literal. The delay before and after each fork consists of $9C+3V$ single-use paths. The right end of the last clause is connected to a White goal location.

A race gadget is connected to the right end of the last variable gadget, with the goal locations such that Player 1 is the player with a win gadget inside the last variable gadget. The path with a White win gadget, which Black will walk down, is followed by $C(18C+6V+1)+2$ of single-use paths in series leading to a Black win gadget. The other path, which White will walk down, is connected to the first clause gadget.

Both players begin at the left end of the first variable gadget, and White goes first.

The game begins with White choosing a branch of the first variable gadget, corresponding to a choice of variable, and Black taking the other branch. Then Black chooses a branch of the second variable gadget, choosing the assignment of the variable based on the path White is forced to take. The players continue to take turns assigning variables. If either player deviates from this, such as by going into the delay in a clause gadget or by going backwards along another path, the other player will reach the race gadget first and win; the delay in clause gadgets is long enough to ensure that they do not have time to get through the clause gadget before losing. Otherwise both players arrive at the race gadget, and are sent down different branches.

White then proceed through each clause in series. Each branch of a clause is traversable if and only if the corresponding literal is true (since White took the bottom side and Black took the top side of each clause). The single-use paths between forks ensure that White cannot do anything other than progress through each clause gadget. If the formula is satisfied, White has a path through the clauses, and wins after $C(18C+6V+1)$ turns. If the formula is not satisfied, Black, who is walking down their long path, wins after slightly longer. Thus White has a forced win if and only if the quantified formula is true.
\end{proof}

\begin{lemma}\label{lem:2-player bounded one-way}
2-player motion planning with the single-use one-way gadget is PSPACE-complete.
\end{lemma}
\begin{proof}
We again reduce from QBF. In the reduction in Lemma~\ref{lem:2-player bounded two-way}, neither player ever has to move through a single-use gadget to the left. Thus we can replace each bidirectional single-use gadget with a one-way single-use gadget pointing to the right, and the reduction still works.
\end{proof}

\begin{corollary}\label{cor:2-player bounded pspace-complete}
2-player motion planning with any nontrivial DAG gadget is PSPACE-complete.
\end{corollary}
\begin{proof}
As noted in Section~\ref{sec:1-Player Bounded} all DAG gadgets contain a single-use transition. This can be bidirectional or one-way, which are both shown to be PSPACE-hard in Lemmas~\ref{lem:2-player bounded two-way} and \ref{lem:2-player bounded one-way}. Containment in PSPACE is given by Lemma~\ref{lem:2-player bounded in pspace}. 
\end{proof}

\section{Team Bounded Motion Planning}
\label{sec:Team Bounded}
In this section we characterize the complexity of team imperfect information motion planning games with DAG gadgets. Since DAG gadgets are inherently bounded, the problem is in NEXPTIME, shown in Lemma~\ref{lem:team bounded in nexp}. We go on to show in Lemma~\ref{lem:team bounded two-way} that any nontrivial DAG gadget is NEXPTIME-complete by first giving a reduction from dependency quantified boolean formula (DQBF) for the single-use gadget. We then show that this proof adapts for single-use one-way gadgets. Since all DAG gadgets with at least one transition contain at least one of these, we achieve hardness for all such DAG gadgets.

\begin{lemma}\label{lem:team bounded in nexp}
Team motion planning with any set of DAG gadgets is in NEXPTIME.
\end{lemma}

\begin{proof}
A \emph{partial history} for a player is the sequence of visible gadget states and moves made by that player, up to some point in the game. A \emph{strategy} is a family of functions, one for each White player, that assign to each possible partial history a legal move from the position at the end of the partial history.

Since the gadget is a DAG, the game lasts a polynomial number of turns. Each player has polynomially many choices for each move, so there are only exponentially many possible sequences of moves, and only exponentially many possible partial histories for each player. Thus a strategy can be written in an exponential amount of space.

To determine whether White has a forced win in the team game, first nondeterministically pick a strategy. Then, for each possible sequence of moves the Black players could make, simulate the game with the White players following the strategy. If Black ever wins, reject; if White always wins, accept. This nondeterministic algorithm accepts if and only if there is some strategy White can use to force a win. The algorithm runs in exponential time because there are exponentially many sequences of moves the Black players might make, and the game for each such sequence takes a polynomial amount of time to simulate. Thus the algorithm decides the team game on systems of the gadget in NEXPTIME.
\end{proof}

\begin{lemma}\label{lem:team bounded two-way}
Team motion planning with the single-use bidirectional gadget is NEXPTIME-complete.
\end{lemma}

\begin{figure}
  \centering
  \includegraphics[scale=0.6]{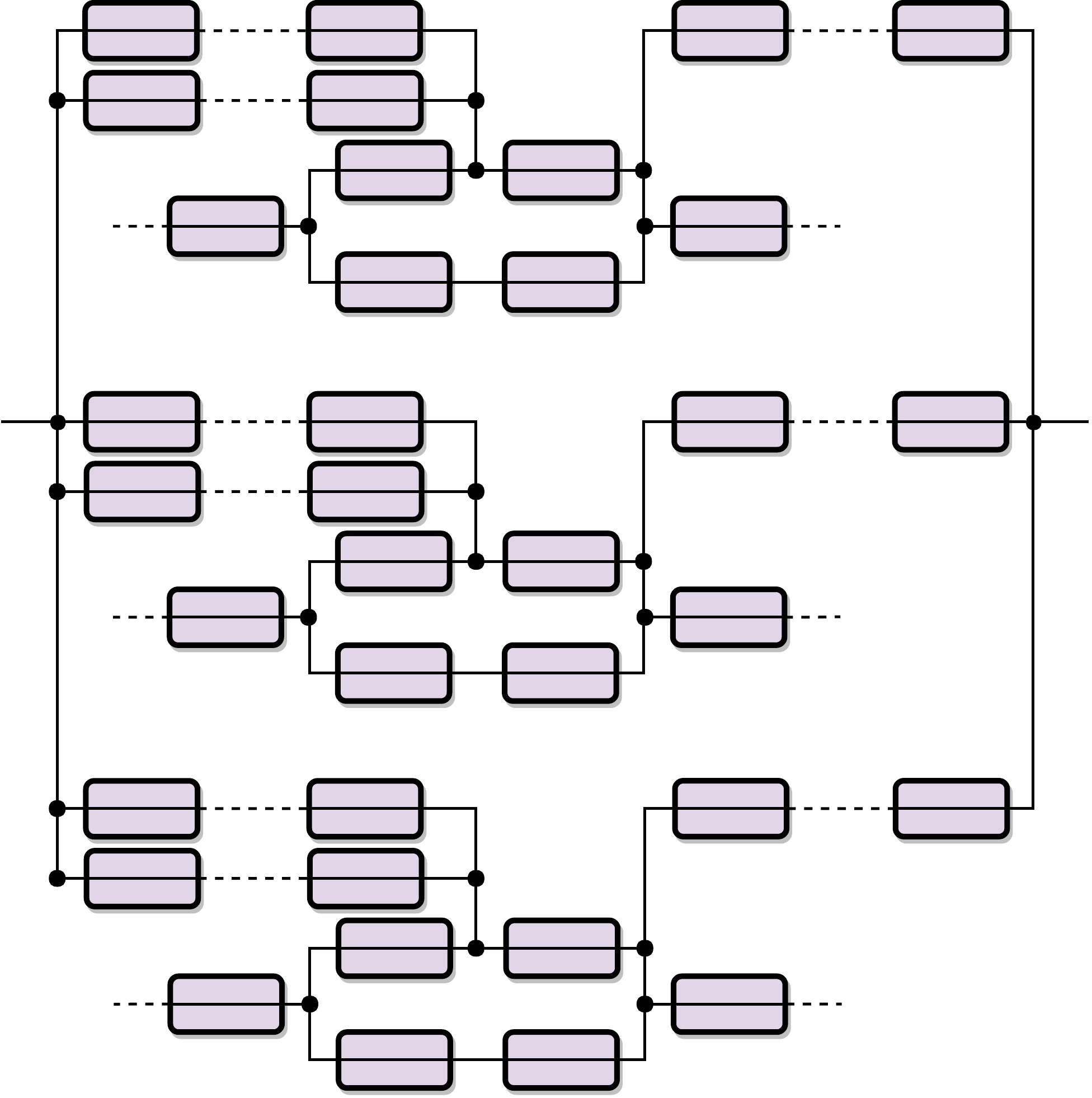}
  \caption{A clause gadget for team games. There are now two paths from the entrance of the clause to each fork, so the White player traversing the clause can return if they discover the fork is not traversable.}
  \label{fig:team bounded clause}
\end{figure}

\begin{figure}
  \centering
  \includegraphics[width=\linewidth]{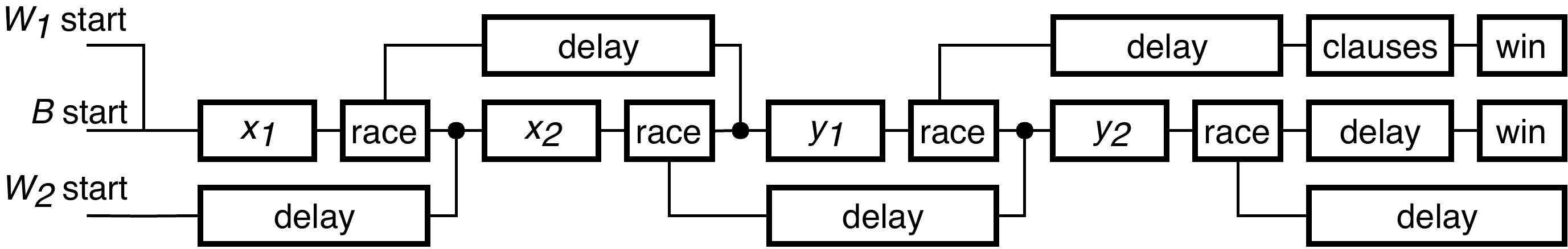}
  \caption{The high-level structure of the DQBF reduction.}
  \label{fig:team bounded schematic}
\end{figure}

\begin{proof}
Containment in NEXPTIME follows from Lemma~\ref{lem:team bounded in nexp}. For NEXPTIME-completeness, we reduce from dependency quantified boolean formulas (DQBF). See Appendix~\ref{app:formula games} for a definition of DQBF. In this reduction White represents the existential variables and Black represents the universal variables.

The reduction uses the same gadgets as that in Lemma~\ref{lem:2-player bounded two-way}, except that the clause gadget is modified as in Figure~\ref{fig:team bounded clause}. This allows the White player checking the formula to try each literal, and return to the start of the clause gadget if the literal is false. This is necessary because the White player cannot see the state of the literals until arriving at them. For variable gadgets, we do not include the portion with a win gadget for Player 2 (the rightmost quarter or so in Figure~\ref{fig:2p bounded var}), since we no longer want players to alternate choosing variables.

We construct the system of gadgets as follows. The overall structure is shown in Figure~\ref{fig:team bounded schematic}. For each set of variables $\vec x_1$, $\vec x_1$, $\vec y_1$, and $\vec y_2$, there is a corresponding set of variable gadgets (without the win gadget component) connected in series, followed by a race gadget. For simplicity, we will put $C$ forks in each branch of each variable, where the formula has $C$ clauses, though usually we need much fewer. Then each variable gadget takes $k=3C+1$ turns to traverse. We call the top path of a race gadget the \emph{fast exit} and the bottom path the \emph{slow exit}, since (in normal play) the first (second) player to arrive leaves through the fast (slow) exit. It will become clear which player each win gadget in a race gadget is for.

The turn order will be $B$, then $W_1$, then $W_2$. Both $B$ and $W_1$ start at the beginning of the variable gadgets for $\vec x_1$. $W_2$ starts next to a delay line of length $d_1$. The fast exit of the race gadget for $\vec x_1$ and the end of this delay line both connect to the beginning of the $\vec x_2$ variable gadgets. The slow exit connects to a delay line of length $d_2$. The end of this delay line and the fast exit of the $\vec x_2$ race gadget connect to the beginning of the $\vec y_1$ variable gadgets, and the slow exit connects to a delay line of length $d_3$. The end of this delay line is connected to the \textit{slow} exit of the $\vec y_1$ race gadget and the beginning of the $\vec y_2$ variable gadgets. The fast exit of the $\vec y_1$ race gadget is connected to yet another delay line of length $d_4$. The slow exit of the $\vec y_2$ race gadget is connected to a long delay line of length $d_5$ followed by a win gadget for $B$, and the fast exit is connected to a longer delay line of length $d_5+3$.

This all serves to accomplish the following. First, $B$ chooses the assignment for $\vec x_1$ accompanied by $W_1$, so $W_1$ learns the assignment. Then $B$ and $W_1$ are separated, and $B$ assigns $\vec x_2$ accompanied by $W_2$. Next, $W_1$ chooses $\vec y_1$ accompanied by $B$, and finally $W_2$ chooses $\vec y_2$ accompanied by $B$. The delays $d_1$ through $d_4$ are chosen so that the White players arrive at exactly the right time; we have $d_1=|\vec x_1|k+1$, $d_2=|\vec x_2|k-1$, $d_3=|\vec y_1|k$, and $d_4=|\vec y_2|$. If a player deviates during variable assignment, they will arrive at their next race gadget too late, and lose.

The end of the final delay line for $W_1$, of length $d_4$, is connected to the first clause gadget, and the clause gadgets are connected in series corresponding to the clauses of the formula. The delay lines in each branch of each clause gadget have length $Vk$, where $V$ is the number of variables; this ensures that if a player enters one of the delay lines during variable selection, an opponent will reach a race gadget and win before they accomplish anything. The end of the last clause gadget is connected to a win gadget for $W_1$. When $W_1$ reaches each clause gadget, they try the literals one at a time. When they cross the delay line to the fork, if the fork is traversable, they move on to the next clause. Otherwise they return through the other delay line and try the next literal. Each clause takes up to $6Vk+1$ turns to cross.

If the formula is satisfied, $W_1$ eventually gets through all the clauses and wins. Otherwise, $B$ wins after walking through their delay line of length $d_5$, which we can set to $C(6Vk+1)+1$.

We have seen that no player or team can benefit by deviating from normal play, and normal play is equivalent to the game corresponding to the DQBF. Thus White has a forced win if and only if the DQBF is true.
\end{proof}

\begin{lemma}\label{lem:team bounded one-way}
Team motion planning with the single-use one-way gadget is NEXPTIME-complete.
\end{lemma}

\begin{proof}
The reduction in Lemma~\ref{lem:team bounded two-way} still works when we replace each single-use bidirectional gadget with a one-way bidirectional gadget. We have to be a bit more careful than in Lemma~\ref{lem:2-player bounded one-way}: of the two paths in a clause gadget from the beginning to a fork, we need one path to point to the right and the other to point to the left, allowing $W_1$ to return from that fork. All other gadgets point to the right.
\end{proof}

\begin{corollary}\label{cor:team bounded nexp-complete}
Team motion planning with any nontrivial DAG gadget is NEXPTIME-complete.
\end{corollary}

\begin{proof}
Every DAG gadget has a single-use transition, which may be either bidirectional or one-way. Both cases are shown to be NEXPTIME-hard in Lemmas~\ref{lem:team bounded two-way} and \ref{lem:team bounded one-way}. Containment in NEXPTIME is Lemma~\ref{lem:team bounded in nexp}.
\end{proof}

\section{Applications}\label{sec:applications}
In this section we give examples of some known hard problems whose proofs can be simplified by using this motion planning framework.

\subsection{PushPull-1F}
\label{sec:application pushpull}

In this section, we use the results of this paper to provide a simple proof that a Sokoban variant called PushPull-$1$F is PSPACE-hard, by reducing from motion planning in planar systems of locking 2-toggles (Section~\ref{sec:1-Player Unbounded Planar}). This problem, and many related problems, were considered in \cite{us} and were shown to be PSPACE-complete in \cite{Pereira-Ritt-Buriol-2016} by a reduction from nondeterministic constraint logic; our reduction is much more straightforward using the infrastructure of the gadget framework.

\begin{definition}
  In \emph{PushPull-$1$F}, there is a square grid containing movable blocks, fixed blocks, an agent, and a goal location. The agent can freely move through empty squares, but can't move through blocks. The agent can push or pull one movable block at a time. The agent wins by reaching the goal location. The corresponding decision problem is whether a given instance of PushPull-$1$F is winnable.
\end{definition}

In the notation `PushPull-$1$F,' `PushPull' indicates that the agent can both push and pull, `$1$' indicates the number of blocks which can be moved at a time, and `F' indicates the existence of fixed blocks \cite{us}. 

\begin{theorem}[\cite{Pereira-Ritt-Buriol-2016}]
  PushPull-$k$F is PSPACE-hard for $k\geq1$.
\end{theorem}

\begin{proof}
  We reduce from 1-player planar motion planning with locking 2-toggles, shown PSPACE-complete in Theorem~\ref{thm:planar 1p unbounded}. The (planar) connection graph is implemented using tunnels built with fixed blocks, and the agent and target location are placed appropriately. It suffices to build a gadget which behaves as a locking 2-toggle.

  Such a gadget is shown in Figure~\ref{fig:pushpull-1f l2t}. The two tunnels, currently both traversable, go from top to left and right to bottom. They interact in the center, where traversing either tunnel requires pushing a block into the middle square, which blocks the other tunnel. This is surrounded by four 1-toggles, which prevent additional traversals which aren't possible in a locking 2-toggle. Each 1-toggle is a room with 3 blocks, which can only be entered on one side. Upon entry, the agent can move the blocks to reveal the other exit, but doing so requires blocking the entrance taken, which flips the 1-toggle.

  \begin{figure}
    \centering
    \includegraphics[width=.8\linewidth]{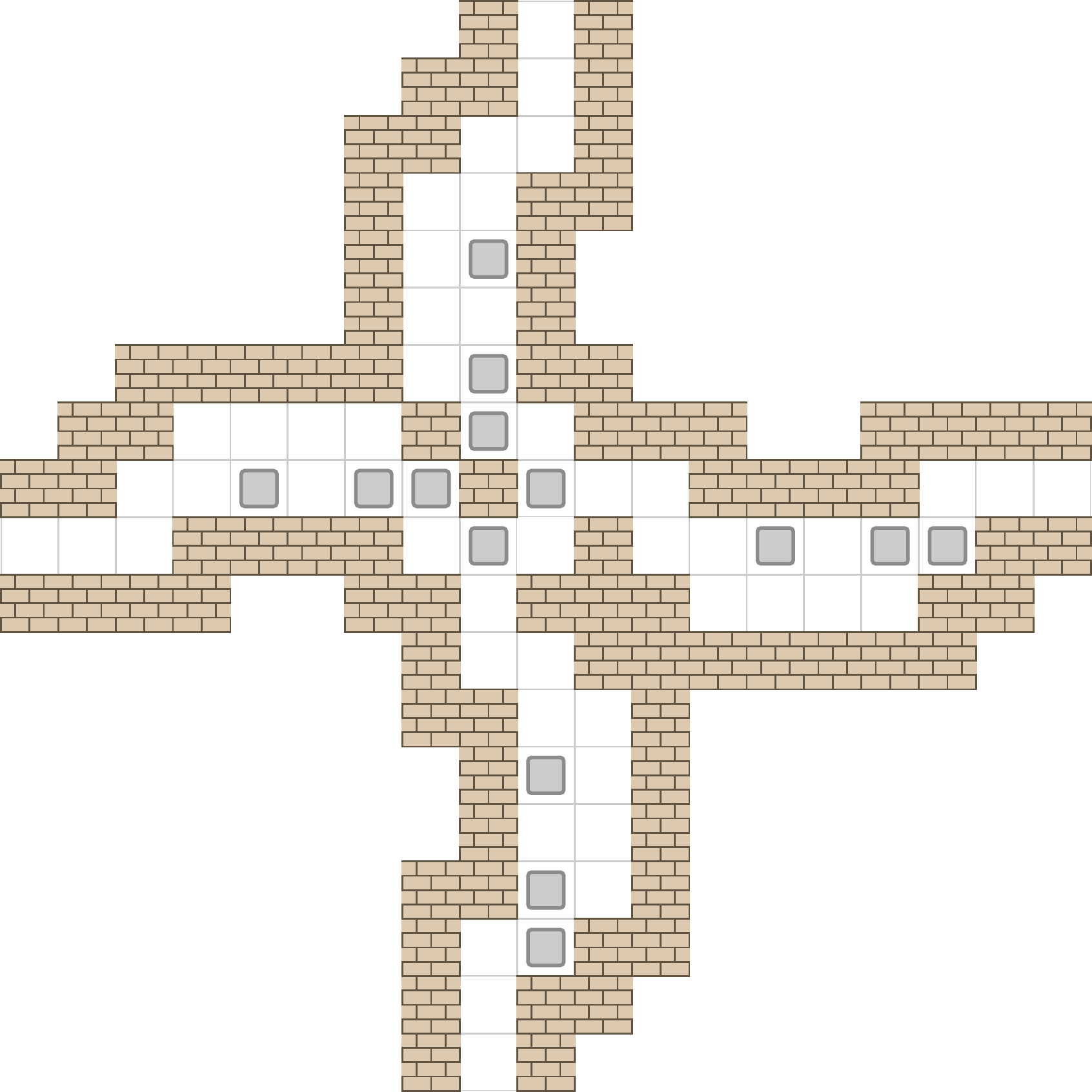}
    \caption{A locking 2-toggle in PushPull-1F.}
    \label{fig:pushpull-1f l2t}
  \end{figure}  
\end{proof}

\subsection{Mario Kart}
Mario Kart is a popular Nintendo racing game whose computational complexity was considered in \cite{MarioKart} which showed NP-completeness for 1 player races and PSPACE-completeness for 2 player races with reductions from 3SAT and QSAT respectively. Using results from this paper, the 2 player proof now only needs a single, simple gadget, reducing a several page proof to a paragraph.

\begin{theorem}
Deciding if a player can force a win in two player Mario Kart is PSPACE-hard.
\end{theorem}
\begin{proof}
A single-use one-way gadget can be constructed from a ramp and Dash Mushroom in Mario Kart. We place a ramp before a gap in the track long enough that a racer going at the normal maximum speed will not be able to make the jump and will fall onto another track that will take a long time to reach the finish line, ensuring they lose. However, this gap is small enough that, if the player uses a Dash Mushroom before, the increase in speed will allow them to make the jump. We put a single Dash Mushroom power-up before each ramp, ensuring the first racer to arrive can pick up the item and use it to cross the gap. To ensure a racer does not pick up the item and then keep it for later use, we precede the mushroom and ramp with a one-way gadget implemented by a long-fall. Along with the trivial existence of crossovers and the finish line as a location based win condition, Mario Kart is PSPACE-hard by Theorem~\ref{lem:2-player bounded one-way}.
\end{proof}

\section{Open Problems}
\label{sec:open problems}

This paper characterizes the complexity of two large classes of gadgets
(DAG gadgets and reversible deterministic gadgets).
Ideally, we could fully characterize the complexity of motion planning for
every gadget type (and set of gadgets) as being easy or hard.
There are many specific steps we might take towards this grand goal:

\begin{enumerate}
\item Is 2-player motion planning with 1-toggles EXPTIME-complete?
  This would complete our characterization for 2-player games
  with $k$-tunnel reversible deterministic gadgets.
  As an easier target, we could prove PSPACE-hardness, perhaps
  by adapting the 2-player proof for one-way closing gadgets.
\item Can we extend our characterizations of $k$-tunnel reversible
  deterministic gadgets to remove one of these restrictions?  Specifically,
  non-tunnel gadgets, non-reversible gadgets, and nondeterministic gadgets
  are all interesting (and challenging) goals.
\item Which motion planning problems remain hard on planar systems of gadgets,
  like we proved for 1-player reversible deterministic?
  Are there any examples of gadgets where the planar version of the
  motion planning problem has a different complexity?

\end{enumerate}

While we focused in this paper on general theory building, we can also explore
the application of this motion planning framework to analyze the complexity of
specific problems of interest. We conjecture that the results of this paper
simplify many past hardness proofs, which can now be reduced to one or two
figures showing how to build any hard gadget according to our
characterization, and how to connect gadgets together. See the hardness
surveys \cite{AlgGameTheory_GONC3,GPCBook09,NPPUzzles08,6.890}
for a large family of candidate problems. Of course, we also
hope that this framework will enable the solution of many open problems in
this space.

\bibliography{bibliography}

\appendix
\section{Problem Definitions}

In this appendix, we give formal definitions for the known hard problems used in this paper.
In the paper we use single player, 2-player, and team imperfect information versions of Constraint Logic and Boolean Formula Games. The exact problems are specified in the following sections.

\subsection{Constraint Logic}
\label{app:constraint logic}

Constraint Logic \cite{CL_Complexity2008,GPCBook09} is a uniform family of
games --- one-player, two-player, or team, with both bounded and unbounded
variants --- with the appropriate complexity in each case (as in
Table~\ref{results}). We will only describe the unbounded variants of
Constraint Logic, as we use formula games for our bounded reductions. We also
do not describe zero-player Constraint Logic, as we do not need it here.

In general, a \emph{constraint graph} is an undirected maximum-degree-3 graph,
where each edge has a weight of $1$ (called a red edge) or $2$ (called a blue
edge). A \emph{legal configuration} of a constraint graph is an orientation of
the edges such that, at every vertex, the total incoming weight is at
least~$2$. A \emph{legal move} in a legal configuration of a constraint graph
is a reversal of a single edge that results in another legal configuration.

In \emph{1-player} Constraint Logic
(also called \emph{Nondeterministic Constraint Logic or NCL}),
we are given a legal configuration of a constraint graph and a target
edge $e$, and we want to know whether there is a sequence of legal moves
ending with the reversal of target edge~$e$. In this game, two types of
vertices suffice for PSPACE-completeness: an \textsc{OR} vertex has
exactly three incident blue edges, and an \textsc{AND} vertex has
exactly one incident blue edge and exactly two incident red edges.
We can also assume that each \textsc{OR} vertex can be assigned two ``input''
edges, and the overall construction is designed to guarantee that at most
one input edge is incoming at any time; thus, we only need a
``Protected \textsc{OR}'' gadget which does not handle the case of two
incoming inputs.
Furthermore, the problem remains PSPACE-complete for planar constraint graphs.

In \emph{2-player Constraint Logic (2CL)},
each edge of a constraint graph is also
colored either black or white, and two players named Black and White alternate
making valid moves where each player can only reverse an edge of their color.
Given a legal configuration of a constraint graph, a target white edge for
White, and a target black edge for Black, the goal is to determine whether
White has a forced win, i.e., a strategy for reversing their target edge
before Black can possibly reverse their target edge.  In this game, six types
of vertices suffice for EXPTIME-completeness: \textsc{and} and \textsc{or}
vertices where all edges are white, \textsc{and} vertices where all edges are
black, \textsc{and} vertices where the blue edge is white and one or both of
the red edges are black, and degree-2 vertices where exactly one edge is
black.

In \emph{Team Private Constraint Logic (TPCL)},
there are two players on the White team and
one player on the Black team, who play in round-robin fashion. In each move,
the player can reverse up to a constant number $k$ of edges of their color.
Each player has a target edge to reverse, and can see the orientation of a specified set of edges,
including edges of their own color and edges incident to those edges. Given a legal
configuration of a constraint graph, the goal is to determine whether the
White team has a forced win; i.e., whether one of the White players can reverse their
target edge before Black can. In this game, all possible black/white colorings
of \textsc{and} and \textsc{or} vertices suffice for RE-completeness.
(Only undecidability has been claimed before, but RE-completeness follows
by the same arguments.)

\subsection{Formula Games}\label{app:formula games}

A \emph{3-CNF formula} is a boolean formula $\varphi$ of the form $C_1\wedge\dots\wedge C_k$, where each \emph{clause} $C_i$ is the disjunction of up to three \emph{literals}, which are variables or their negations. An \emph{assignment} for such a formula specifies a truth value for each variable, and is \emph{satisfying} if the formula is true under the assignment.\

In \emph{3SAT}, we are given a 3-CNF formula, and we want to know whether it has a satisfying assignment. 3SAT is NP-complete \cite{NPBook}.

A \emph{partially quantified boolean formula} is a formula of the form $Q_1x_1: \cdots : Q_nx_n : \varphi$, where $Q_i$ is one of the quantifiers $\forall$ or $\exists$, $x_i$ is a (distinct) variable, and $\varphi$ is a 3-CNF formula. An \emph{assignment} for a partially quantified boolean formula specifies a truth value for each variable in $\varphi$ that is not any $x_i$, called \emph{free} variables. For a partially quantified boolean formula $\psi=Q_1x_1 : \cdots : Q_nx_n : \varphi$ with $n>0$, let $\psi^\prime=Q_2x_2 : \cdots : Q_nx_n : \varphi$. Given an assignment $S$ for $\psi$, define assignments $S+x_1$ and $S+\neg x_1$ for $\psi^\prime$ which assign the same truth value as $S$ to each free variable of $\varphi$ and assign `true' and `false' to $x_1$, respectively. The truth value of $\psi$ under $S$ is defined recursively as follows:
\begin{itemize}
  \item If $n=0$ (so $\psi=\varphi$), $\psi$ is true under $S$ if and only if $\varphi$ is true under $S$.
  \item If $n>0$ and $Q_1=\forall$, $\psi$ is true under $S$ if and only if $\psi^\prime$ is true under both $S+x_1$ and $S+\neg x_1$.
  \item If $n>0$ and $Q_1=\exists$, $\psi$ is true under $S$ if and only if $\psi^\prime$ is true under at least one of $S+x_1$ and $S+\neg x_1$.
\end{itemize}

A \emph{quantified boolean formula} is a partially quantified boolean formula with no free variables. A quantified boolean formula has only one assignment (which is empty), so we say it is true if it is true under this unique assignment. 

The truth value of a quantified boolean formula $\psi=Q_1x_1: \cdots : Q_nx_n: \varphi$ is equivalent to whether $\exists$ has a forced win in the following game: two players $\exists$ and $\forall$ choose an assignment for $\varphi$ by assigning variables in the order they are quantified, with player $Q_i$ choosing the truth value of $x_i$. $\exists$ wins if the assignment satisfies $\varphi$.

In \emph{QBF}, we are given a (fully) quantified boolean formula, and we want to know whether it is true. QBF is PSPACE-complete, even if we restrict to formulas with alternating quantifiers beginning with $\exists$. This restriction is equivalent to that $\exists$ and $\forall$ take alternating turns, with $\exists$ going first \cite{NPBook}.

A \emph{dependency quantified boolean formula} is a formula of the form $\forall x_1 : \cdots : \forall x_m : \exists y_1(s_1) : \cdots : \exists y_n(s_n) : \varphi$, where $x_i$ and $y_j$ are (distinct) variables, $\varphi$ is a 3-CNF formula, and $s_j$ is a subset of $\{x_i\mid i\le m\}$. We also require that every variable in $\varphi$ is some $x_i$ or $y_j$ ($\varphi$ has no free variables). A \emph{strategy} for a dependency quantified boolean formula is a collection of functions $f_j:\{\text{true},\text{false}\}^{s_j}\to\{\text{true},\text{false}\}$ for $j=1,\dots,n$. A strategy \emph{solves} a dependency quantified boolean formula if for every map $S:\{x_i\mid i\le m\}\to\{\text{true},\text{false}\}$, the assignment given by $x_i\mapsto S(x_i)$ and $y_j\mapsto f_j(S|_{s_j})$ satisfies $\varphi$. Intuitively, $y_j$ is only allowed to depend on the variables in $s_j$. A quantified boolean formula is a special case of a dependency quantified boolean formula, where each $s_j=\{x_i\mid i<k\}$ for some $k$. A dependency quantified boolean formula is \emph{true} if there is a strategy that solves it.

The truth value of a dependency quantified boolean formula $\forall x_1 : \cdots : \forall x_m : \exists y_1(s_1) : \cdots : \exists y_n(s_n) : \varphi$ is equivalent to whether the $\exists$ team has a forced win in the following game, which puts a team of one player $\forall$ against a team of players $\exists_j$ for $j=1,\dots,n$: $\forall$ picks a truth value for each $x_i$. $\exists_j$ sees the truth value for each element of $s_j$ (and nothing else) and picks a truth value for $y_j$. The $\exists$ team wins if the resulting assignment satisfies $\varphi$.

In the \emph{DQBF} problem, we are given a dependency quantified boolean formula, and we want to know whether it is true. DQBF is NEXPTIME-complete even if we restrict to formulas of the form $\forall\vec x_1 : \forall \vec x_2 : \exists\vec y_1(\vec x_1) : \exists \vec y_2(\vec x_2) : \varphi$, where $\vec x_i$ and $\vec y_i$ may contain multiple variables, and each variable in $\vec y_i$ can depend on all the variables in $\vec x_i$. This restriction is equivalent to requiring that the $\exists$ team has two players who each choose multiple variables, and they see disjoint exhaustive subsets of the variables $\forall$ picks \cite{Peterson:1979:MA:1382433.1382634}.

\end{document}